\documentclass[twocolumn,superscriptaddress,longbibliography,prd]{revtex4-2}
\usepackage{times}
\usepackage[T1]{fontenc}

\usepackage{physics}
\usepackage{amsmath,amssymb,mathrsfs}
\usepackage[dvipdfm]{graphicx}
\usepackage{dcolumn}
\usepackage{bm}
\usepackage{morefloats}
\usepackage{multirow}
\usepackage{amssymb}
\usepackage{makecell}
\usepackage{amsmath}
\usepackage{xcolor}
\usepackage{longtable}
\usepackage{fix-cm}
\usepackage{verbatim}
\usepackage{float}
\usepackage[T1]{fontenc}
\usepackage{hyperref}
\hypersetup{colorlinks=true, allcolors=blue}
\usepackage{booktabs}
\usepackage{mathtools}
\usepackage[dvipsnames]{xcolor}

\makeatletter
\def\NAT@def@citea{\def\@citea{\NAT@separator}}
\makeatother

\def\0\\{\nonumber\\}

\newcommand{\beq}{\begin{equation}}
\newcommand{\eeq}{\end{equation}}
\newcommand{\beqn}{\begin{eqnarray}}
\newcommand{\eeqn}{\end{eqnarray}}

\makeatletter
\newcommand\footnoteref[1]{\protected@xdef\@thefnmark{\ref{#1}}\@footnotemark}
\makeatother

\begin{document}

%\preprint{APS/123-QED}

\title{
Effects of Differential Rotation on the Maximum Mass of Neutron Stars
}

\author{Hyukjin Kwon}
\email{kwon.h.04c4@m.isct.ac.jp}
\affiliation{Department of Physics, School of Science, Institute of Science Tokyo, Tokyo 152-8550, Japan}

\author{Jinho Kim}
\email{jkim@kasi.re.kr}
\affiliation{Korea Astronomy and Space Science Institute, 776 Daedeok-daero, Yuseong-gu, Daejeon 34055, South Korea}

\author{Kazuyuki Sekizawa}
\email{sekizawa@phys.sci.isct.ac.jp}
\affiliation{Department of Physics, School of Science, Institute of Science Tokyo, Tokyo 152-8550, Japan}
\affiliation{Nuclear Physics Division, Center for Computational Sciences, University of Tsukuba, Ibaraki 305-8577, Japan}
\affiliation{RIKEN Nishina Center, Saitama 351-0198, Japan}

\date{\today}% It is always \today, today,

\begin{abstract}
The maximum mass of neutron stars provides a key constraint on the
equation of state (EoS) of dense matter. Recent observations, including
the ${\approx}2\,M_{\odot}$ pulsar PSR~J0740+6620, have placed strong constraints on a
large class of soft EoSs, while the possible existence of a compact
object with a mass of $2.50$--$2.67$\,$M_{\odot}$ in GW\,190814 further
challenges our understanding of dense matter. Moreover, the inclusion of
hyperonic degrees of freedom generally softens the EoS, making it
difficult to support massive neutron stars even when the $2$\,$M_{\odot}$
constraint is satisfied (a problem known as the hyperon puzzle). In this work, we investigate whether differential rotation can enhance the maximum mass of neutron stars constructed with an EoS including hyperons, thereby addressing the maximum-mass constraints
imposed by current observations. We employ the Cook-Shapiro-Teukolsky
(CST) approach, a numerically improved reformulation of the
Komatsu-Eriguchi-Hachisu (KEH) scheme, to construct equilibrium
configurations of differentially rotating neutron stars. For the nuclear matter
EoS, we adopt a relativistic mean-field (RMF) model incorporating
hyperonic degrees of freedom through an SU(6) symmetric coupling scheme.
We find that differential rotation can substantially increase the maximum mass, yielding configurations consistent with the mass range inferred from GW\,190814. However, a sufficiently soft EoS fails to satisfy the constraint from PSR~J0740+6620 (346\,Hz) even with differential rotation applied. We also present a systematic analysis of the internal structure of the resulting equilibrium configurations. Furthermore, we demonstrate the existence of quasi-toroidal configurations and present equilibrium sequences incorporating the full baryon octet under extreme differential rotation.
\end{abstract}
 
\maketitle

\section{Introduction}\label{Sec:Intro}

Theoretical descriptions of nuclear many-body systems have been developed to reproduce experimental observables measured in terrestrial laboratories.
Key benchmarks include the binding energies and charge radii of finite
nuclei~\cite{Angeli2013,Wang2021}, as well as bulk properties of nuclear
matter such as the symmetry energy~\cite{LI2013} and
incompressibility~\cite{Shlomo2006}, which are constrained by collective
excitations. However, such constraints obtained from terrestrial experiments
are largely limited to densities around nuclear saturation density.

In contrast, neutron stars span a wide range of densities, from
sub-saturation densities in the crust to several times saturation density
in their cores~\cite{Burgio2020}. They therefore provide a unique
astrophysical laboratory for probing the equation of state (EoS) of dense
nuclear matter. By anchoring nuclear theories at saturation density and extending
them to higher densities, neutron star observations enable a systematic
extrapolation and refinement of the EoS beyond the regime accessible via
terrestrial experiments.

The masses of most known neutron stars have been precisely determined through general relativistic effects in binary systems, in particular the Shapiro delay. More recently, gravitational-wave observations have begun to reveal more massive—and in some cases ambiguous—compact objects.
Among various neutron star observables, the maximum mass provides one of
the most stringent constraints on the high-density behavior of the EoS.
Observations of PSR~J0348+0432 (25.56\,Hz)~\cite{Antoniadis2013},
PSR~J1614$-$2230 (317\,Hz)~\cite{Fonseca2016}, and
PSR~J0740+6620 (346\,Hz)~\cite{Cromartie2020,Fonseca2021} have confirmed
the existence of neutron stars with masses of approximately
$2\,M_{\odot}$, excluding a large class of soft EoSs derived from
conventional nuclear interactions.

The nature of the secondary component in the gravitational-wave event
GW\,190814~\cite{Abbott2020} further motivates the study of massive neutron
stars. The measured mass of this object, $2.50$--$2.67\,M_{\odot}$, lies
in the so-called mass gap between the maximum neutron star mass and the
minimum black hole mass. Whether this object is a heavy neutron star or a
light black hole remains an open question.

Even when an EoS satisfies the $2\,M_{\odot}$ constraint, an additional difficulty arises with the inclusion of hyperonic degrees of freedom. The appearance of hyperons generally softens the EoS, making it difficult to support such massive neutron stars~\cite{Glendenning1985}. This issue,
known as the hyperon puzzle, remains one of the central open problems in
nuclear and hadron physics~\cite{Chatterjee2016}. Various approaches have
been proposed to address this challenge, including reparameterizing the
EoS~\cite{Chen2014,Gonzalez2018}, introducing additional repulsive
interactions or three-body forces in the hyperon
sector~\cite{Weissenborn2012,Drago2016,Vidana2011,Yamamoto2013,Lonardoni2015,Lee2026},
considering phase transitions to deconfined quark
matter~\cite{Fujimoto2026}, and incorporating light fermionic dark
matter~\cite{Li2012,Guha2024}.

From an astrophysical perspective, rotation provides an additional
mechanism for supporting larger neutron star masses. For rotating neutron star structures, two well-known approaches exist: the first one is the Komatsu-Eriguchi-Hachisu (KEH) method~\cite{Komatsu1989a,Komatsu1989b} which solves the integral form of the Einstein equations, and the second one is the Bonazzola-Gourgoulhon-Salgado-Marck (BGSM) method, which solves elliptic equations using a spectral method~\cite{Bonazzola1993,Bonazzola1998}. These two methods have been shown to be in good agreement with each other~\cite{Nozawa1998}. Kwon and Sekizawa investigated whether the spin frequency of PSR~J0740+6620 can affect the maximum mass using the KEH method, finding only a marginal increase due to rigid rotation~\cite{Kwon2026}. Similarly, even for the fastest known pulsar, PSR~J1748$-$2446ad (716\,Hz)~\cite{Hessels2006}, rigid rotation alone is
insufficient to support masses as large as $2.5\,M_{\odot}$. Zhang
\textit{et al.} examined this possibility using the \texttt{RNS}
code~\cite{Stergioulas1995} and concluded that neutron stars can sustain such masses only near the Keplerian limit~\cite{Zhang2020}.

In contrast, differential rotation can support significantly larger
masses. This was originally discussed in the context of binary neutron
star merger remnants: Shibata and Uryu performed general relativistic
simulations and found that merger remnants can reach masses close to
$3\,M_{\odot}$~\cite{Shibata2000}, objects now referred to as
hypermassive neutron stars (HMNSs). Subsequently, Baumgarte \textit{et
al.} demonstrated that equilibrium configurations exceeding
$3\,M_{\odot}$ can be constructed for differentially rotating neutron
stars within a polytropic EoS~\cite{Baumgarte2000} and several selected realistic EoSs~\cite{Morrison2004} using the KEH
framework.

In this work, we employ the Cook-Shapiro-Teukolsky (CST)
approach~\cite{Cook1992,Cook1994}, a reformulation of the KEH method based on a
compactified coordinate transformation that maps spatial infinity onto a
finite computational domain, thereby allowing asymptotic boundary
conditions to be imposed exactly. This leads to improved numerical
accuracy and convergence compared to the original KEH scheme. We
implement differential rotation within this framework and pursue two main
objectives. First, we examine whether differential rotation at the observed spin frequency of PSR~J0740+6620 can support its measured mass of $2.14^{+0.10}_{-0.09}\,M_{\odot}$ with an EoS including hyperons. We focus on PSR~J0740+6620 rather than PSR~J0348+0432, as the latter is a slowly rotating system whose mass
has recently been revised to ${\approx}1.8\,M_{\odot}$~\cite{Saffer2025}. Second, we
investigate whether differentially rotating neutron stars that allow existence of hyperons can reach massive configurations of $2.5\,M_{\odot}$ or even $3\,M_{\odot}$, motivated respectively by the secondary component of GW\,190814 and by HMNS remnant scenarios in binary neutron star mergers. Furthermore, we examine neutron star structure under extreme conditions, finding quasi-toroidal configurations, and investigate their possible internal structure arising under differential rotation. For the nuclear EoS, we adopt a relativistic mean-field (RMF) framework incorporating hyperonic degrees of freedom through an SU(6) symmetric coupling scheme.

The remainder of this paper is organized as follows. In
Sec.~\ref{Sec:Methods}, we describe the theoretical framework underlying the RMF model, the neutron star EoS, and the KEH method for differentially rotating neutron stars. Numerical procedures and computational details are discussed in Sec.~\ref{Sec:Details}. Results and analysis are presented in Sec.~\ref{Sec:Results}. Summary and prospect are provided in Sec.~\ref{Sec:Conclusion}.

\section{FORMULATION} \label{Sec:Methods}

\subsection{Relativistic Mean Field Theory}

The RMF model was originally proposed by
Walecka~\cite{Walecka1974,Serot1984}, in which the nuclear force is described through the exchange of mesons. In this work, we employ an
extended RMF model incorporating the isoscalar-scalar $\sigma$,
isoscalar-vector $\omega$, and isovector-vector $\vec{\rho}$ mesons, as
well as the strange isoscalar-scalar $\sigma^*$ and isoscalar-vector
$\phi$ mesons. The strange mesons $\sigma^*$ and $\phi$ mediate the
hyperon-hyperon interaction and are essential for a realistic description
of the hyperonic sector~\cite{Schaffner1994}. The Lagrangian density
with hyperonic degrees of freedom is given by:

\begin{widetext}
\begin{equation}
\begin{aligned}
    \mathcal{L} &= \sum_{B} \bar{\psi}_B \bigg[ i\gamma_\mu \partial^\mu
    - M^*_B(\sigma, \sigma^*) - g_{\omega B}\gamma_\mu \omega^\mu
    - g_{\phi B} \gamma_\mu \phi^\mu
    - g_{\rho B} \gamma_\mu \vec{\rho}^\mu \cdot \vec{I}_B \bigg] \psi_B \\
    &\quad + \frac{1}{2} (\partial_\mu \sigma \partial^\mu \sigma
    - m^2_\sigma \sigma^2)
    + \frac{1}{2}(\partial_\mu \sigma^* \partial^\mu \sigma^*
    - m^2_{\sigma^*} {\sigma^*}^2)
    + \frac{1}{2}m^2_\omega\omega_\mu\omega^\mu
    - \frac{1}{4}W_{\mu \nu} W^{\mu \nu} \\
    &\quad + \frac{1}{2}m^2_\phi\phi_\mu\phi^\mu
    - \frac{1}{4}P_{\mu \nu} P^{\mu \nu}
    + \frac{1}{2}m^2_\rho \vec{\rho}_\mu\vec{\rho}^\mu
    - \frac{1}{4}\vec{R}_{\mu \nu} \vec{R}^{\mu \nu}
    - U_\text{NL} (\sigma, \omega^\mu, \vec{\rho}^\mu),
\end{aligned}
\label{LagrangianDensity}
\end{equation}
\end{widetext}
where
\begin{subequations}
\begin{align}
    W_{\mu \nu}       &=\; \partial_\mu \omega_\nu - \partial_\nu \omega_\mu, \\
    P_{\mu \nu}       &=\; \partial_\mu \phi_\nu - \partial_\nu \phi_\mu, \\
    \vec{R}_{\mu \nu} &=\; \partial_\mu \vec{\rho}_\nu - \partial_\nu \vec{\rho}_\mu,
\end{align}
\end{subequations}
with $\psi$ the baryon Dirac field and $\vec{I}_B$ the isospin matrix
for baryon $B = N, \Lambda, \Sigma^{+,0,-}$, and $\Xi^{0,-}$.
$M_B^*$ denotes the density-dependent effective baryon mass,
\begin{equation}
    M^*_B(\sigma,\sigma^*) = M_B - g_{\sigma B} \sigma - g_{\sigma^* B} \sigma^*,
\end{equation}
and the nonlinear self-interaction term $U_\text{NL}$ is given by
\begin{equation}
\begin{aligned}
    U_\text{NL}(\sigma, \omega^\mu, \vec{\rho}^\mu)
    &= \frac{1}{3}g_2 \sigma^3 + \frac{1}{4}g_3 \sigma^4
    - \frac{1}{4}c_3 (\omega_\mu \omega^\mu)^2 \\
    &\quad - \Lambda_\nu g^2_{\rho N} g^2_{\omega N}
    (\omega_\mu \omega^\mu)(\vec{\rho}_\mu \vec{\rho}^\mu).
\end{aligned}
\end{equation}

In the RMF approximation, the meson fields are replaced with their
constant mean-field values $\bar{\sigma}$, $\bar{\sigma}^*$,
$\bar{\omega}$, $\bar{\phi}$, and $\bar{\rho}$. The equations of motion
derived from Eq.~\eqref{LagrangianDensity} for uniform matter are:
\begin{subequations}
\begin{align}
    m^2_\sigma \bar{\sigma} + g_2 \bar{\sigma}^2 + g_3 \bar{\sigma}^3
    &= \sum_B g_{\sigma B} \rho^s_B, \\
    m^2_{\sigma^*} \bar{\sigma}^*
    &= \sum_B g_{\sigma^* B}\rho^s_B, \\
    (m^2_\omega + 2\Lambda_\nu g^2_{\rho N} g^2_{\omega N}
    \bar{\rho}^2)\bar{\omega} + c_3 \bar{\omega}^3
    &= \sum_B g_{\omega B} \rho_B, \\
    m^2_{\phi} \bar{\phi}
    &= \sum_B g_{\phi B}\rho_B, \\
    (m^2_\rho + 2\Lambda_\nu g^2_{\rho N} g^2_{\omega N}
    \bar{\omega}^2)\bar{\rho}
    &= \sum_B g_{\rho B} (\vec{I}_B)_3 \rho_B,
\end{align}
\label{meson}
\end{subequations}
where $(\vec{I}_B)_3$ denotes the third component of the isospin
operator. The scalar density $\rho^s_B$ and the baryon number density
$\rho_B$ are given by:
\begin{subequations}
\begin{align}
    \label{scalardensity}
    \rho^s_B &= \frac{1}{\pi^2} \int^{k_{F_B}}_0 \dd k\, k^2
    \frac{M^*_B}{\sqrt{k^2 + {M^*_B}^2}}, \\
    \rho_B &= \frac{1}{\pi^2} \int^{k_{F_B}}_0 \dd k\, k^2
    = \frac{k^3_{F_B}}{3\pi^2}.
\end{align}
\end{subequations}
The meson field equations~\eqref{meson} and the scalar
density~\eqref{scalardensity} are solved self-consistently. The total
energy density and pressure of the system are then given by
\begin{equation}
\begin{aligned}
    \varepsilon_B &= \frac{1}{\pi^2} \sum_{B} \int^{k_{F_B}}_{0}
    \dd k\, k^2 \sqrt{k^2+M^{*2}_B} \\
    &\quad + \frac{1}{2}(m^2_\sigma \bar{\sigma}^2
    + m^2_{\sigma^*} \bar{\sigma}^{*2}
    + m^2_\omega \bar{\omega}^2
    + m^2_\phi \bar{\phi}^2
    + m^2_\rho \bar{\rho}^2) \\
    &\quad + \frac{1}{3}g_2 \bar{\sigma}^3
    + \frac{1}{4}g_3 \bar{\sigma}^4
    + \frac{3}{4} c_3 \bar{\omega}^4
    + 3\Lambda_{\nu} g^2_{\rho N} g^2_{\omega N} \bar{\omega}^2 \bar{\rho}^2,
\end{aligned}
\end{equation}
and
\begin{equation}
\begin{aligned}
    P_B &= \frac{1}{3\pi^2} \sum_{B}\int^{k_{F_B}}_{0}
    \dd k\, \frac{k^4}{\sqrt{k^2+M^{*2}_B}} \\
    &\quad - \frac{1}{2}(m^2_\sigma \bar{\sigma}^2
    + m^2_{\sigma^*} \bar{\sigma}^{*2}
    - m^2_\omega \bar{\omega}^2
    - m^2_\phi \bar{\phi}^2
    - m^2_\rho \bar{\rho}^2) \\
    &\quad - \frac{1}{3}g_2 \bar{\sigma}^3
    - \frac{1}{4}g_3 \bar{\sigma}^4
    + \frac{1}{4} c_3 \bar{\omega}^4
    + \Lambda_{\nu} g^2_{\rho N} g^2_{\omega N} \bar{\omega}^2 \bar{\rho}^2.
\end{aligned}
\end{equation}

In this work, we adopt the FSUGarnet parametrization for the nucleonic
sector~\cite{Chen2015a,Chen2015b}. The FSUGarnet model was originally
developed without including strange mesons ($\sigma^*$, $\phi$) or
hyperons, but has been shown to simultaneously reproduce the saturation
properties of nuclear matter, the binding energies of closed-shell
nuclei, and neutron star mass-radius constraints consistent with NICER
observations.
 
The saturation density $n_0$ is defined as the density at which the
pressure of symmetric nuclear matter (SNM) vanishes. The key nuclear
matter properties at saturation are:
\begin{eqnarray}
    K_0 &=& 9n_0^2 \left( \frac{\partial^2 E_{\text{SNM}}(n)}{\partial n^2}
    \right)_{n_0} \quad \text{(incompressibility)}, \\
    S_0 &=& \frac{1}{8} \left( \frac{\partial^2 E}{\partial Y_p^2}
    \right)_{Y_p=1/2} \quad \text{(symmetry energy)}, \\
    L &=& 3n_0 \left( \frac{\partial S(n)}{\partial n} \right)_{n_0}
    \quad \text{(slope of $S_0$)},
\end{eqnarray}
where $Y_p$ denotes the proton fraction. The calculated bulk properties
at saturation density are summarized in Table~\ref{Bulk}, and are
consistent with empirical constraints.

%--------------------------------------------------------------------
\begin{table}[t]
\centering
\caption{Bulk properties of nuclear matter at saturation density
calculated with the RMF model with the FSUGarnet parametrization~\cite{Chen2015a,Chen2015b}.}
\begin{tabular}{cccccc}
\hline\hline
$n_0$ (fm$^{-3}$) & $E_0$ (MeV) & $K_0$ (MeV) & $S_0$ (MeV)
& $L$ (MeV) & $M^*_N/M_N$ \\
\hline
0.1532 & $-$16.23 & 229.63 & 30.92 & 50.96 & 0.578 \\
\hline\hline
\end{tabular}
\label{Bulk}
\end{table}
%--------------------------------------------------------------------

The meson-hyperon coupling constants are determined by the SU(6) spin-flavor symmetry relations for the vector mesons~\cite{Schaffner1994,Miyatsu2013}. Although SU(3) flavor symmetry provides a more general framework, it allows a nonzero $\phi$-nucleon coupling $g_{\phi N}$, which modifies the nucleon-nucleon interaction and thereby shifts the nuclear saturation properties. This would require refitting the nucleon sector coupling constants, breaking consistency with the original FSUGarnet parametrization. We therefore adopt the more constrained SU(6) relations, which enforce $g_{\phi N}$ = 0 and allow us to extend the FSUGarnet parameter set to the hyperon sector without any modification of the nucleon sector.

The coupling constants in SU(6) relations are given by
\begin{subequations}
\begin{align}
    \frac{1}{3}g_{\omega N} = \frac{1}{2}g_{\omega \Lambda}
    &= \frac{1}{2}g_{\omega \Sigma} = g_{\omega \Xi}, \\
    2g_{\phi \Lambda} = 2g_{\phi \Sigma}
    &= g_{\phi \Xi} = -\frac{2\sqrt{2}}{3}g_{\omega N}, \\
    g_{\rho N} = \frac{1}{2} &g_{\rho \Sigma} = g_{\rho \Xi}, \\
    g_{\phi N} =& g_{\rho \Lambda} = 0.
\end{align}
\end{subequations}
The scalar meson-hyperon coupling constants $g_{\sigma B}$ and
$g_{\sigma^* B}$ are determined from the hyperon potential depths. In
the RMF approximation, the potential for hyperon $Y$ in SNM reads
\begin{equation}
    U^{(N)}_{Y} = -g_{\sigma Y}\bar{\sigma} + g_{\omega Y} \bar{\omega},
\end{equation}
from which $g_{\sigma Y}$ can be determined. We adopt the following potential
depths at saturation density: $U^{(N)}_{\Lambda} = -27.7$\,MeV,
$U^{(N)}_{\Sigma} = +30$\,MeV, and $U^{(N)}_{\Xi} =
-21$\,MeV~\cite{Batty1997,Kohno2004,Friedman2007,Friedman2023a,Friedman2023b,Friedman2025}.
The coupling constant $g_{\sigma^* B}$ is fixed by the $\Lambda$
potential in $\Lambda$ matter:
\begin{equation}
    U^{(\Lambda)}_{\Lambda} = -g_{\sigma \Lambda}\bar{\sigma}^{(\Lambda)}
    - g_{\sigma^* \Lambda}\bar{\sigma}^{*(\Lambda)}
    + g_{\omega \Lambda} \bar{\omega}^{(\Lambda)}
    + g_{\phi \Lambda} \bar{\phi}^{(\Lambda)}.
\end{equation}
Here, we use the $\Lambda\Lambda$ potential depth 
$U_{\Lambda}^{(\Lambda)} = -5$\,MeV, as inferred from 
the Nagara event~\cite{Takahashi2001}, together with 
the conventional relation $U_\Xi^{(\Xi)} \simeq 
2\,U_\Lambda^{(\Lambda)}$. We use the following values for the masses of the hyperons and strange mesons:
$m_{\sigma^*} = 975\, \text{MeV}$, $m_{\phi} = 1020\, \text{MeV}$, $m_{\Lambda} = 1116\, \text{MeV}$, $m_{\Sigma} = 1193\, \text{MeV}$, and $m_{\Xi} = 1318\, \text{MeV}$. We also assume the coupling constant $g_{\sigma^*N} =0 $. The resulting coupling constants are listed in Table~\ref{tab:coupling}.

%--------------------------------------------------------------------
\begin{table}[t]
\centering
\caption{Hyperon-meson coupling constants used in this work.}
\begin{tabular}{cccccc}
\hline\hline
 & $g_{\sigma B}$ & $g_{\sigma^* B}$ & $g_{\omega B}$
 & $g_{\phi B}$ & $g_{\rho B}$ \\
\hline
$N$      & 10.505 &  0      & 13.700 &  0       &  6.945 \\
$\Lambda$&  6.413 &  5.950  &  9.133 & $-$6.458 &  0     \\
$\Sigma$ &  4.882 &  5.950  &  9.133 & $-$6.458 & 13.890 \\
$\Xi$    &  3.396 & 12.468  &  4.567 & $-$12.917 &  6.945 \\
\hline\hline
\end{tabular}
\label{tab:coupling}
\end{table}
%--------------------------------------------------------------------

\subsection{Neutron star EoS}

To describe the internal structure of neutron stars, we construct the EoS under conditions of charge neutrality and beta equilibrium. The
structure of a neutron star is typically divided into the crust and the core. For the crust, we adopt the Baym--Pethick--Sutherland (BPS)~\cite{Baym1971a} and Baym--Bethe--Pethick (BBP)~\cite{Baym1971b}
EoSs, which are among the most widely used descriptions of the neutron star crust. We set the core-crust transition density at $\rho = 1.5 \times 10^{14} \, \text{g} \, \text{cm}^{-3}$~\cite{Chabanat1997}. For the core region, we construct two EoSs: one consisting
of nucleons and leptons only (nucleonic EoS), and one that additionally includes hyperonic degrees of freedom (hyperonic EoS). These two EoSs are calculated at zero temperature.

\subsubsection{Nucleonic core}

For the nucleonic core, the beta-equilibrium conditions are determined
by the chemical potential equilibria:
\begin{subequations}
\begin{align}
    \mu_n &= \mu_p + \mu_e, \label{eq:betaeq_a} \\
    \mu_e &= \mu_\mu, \label{eq:betaeq_b}
\end{align}
\end{subequations}
together with the charge neutrality condition:
\begin{equation}
    n_p = n_e + n_\mu.
    \label{eq:chargeneutrality_N}
\end{equation}
Muons appear when the electron chemical potential exceeds the muon rest
mass ($m_\mu c^2 \approx 105.7$\,MeV). Tau leptons are not produced in
neutron star matter, as their rest mass ($m_\tau c^2 \approx 1777$\,MeV)
far exceeds the typical lepton chemical potentials in the stellar
interior.

The lepton contribution to the Lagrangian density is:
\begin{equation}
    \mathcal{L}_l = \sum_{l=e,\mu} \bar{\psi}_l
    (i\gamma_\mu \partial^\mu - m_l)\psi_l.
\end{equation}
The lepton energy density and pressure are computed analytically
assuming zero temperature and complete degeneracy as
\begin{subequations}
\begin{align}
    \varepsilon_l &= \frac{m_l^4}{8\pi^2}
    \bigg[ x_l \sqrt{1+x_l^2}(1 + 2x_l^2) - \ln\!\left(x_l + \sqrt{1+x_l^2}\right) \bigg], \\
    P_l &= \frac{m_l^4}{24\pi^2}
    \bigg[ x_l \left(2x_l^2 - 3\right)\sqrt{1+x_l^2} + 3\ln\!\left(x_l + \sqrt{1+x_l^2}\right) \bigg],
\end{align}
\end{subequations}
where $x_l = k_{F,l}/m_l$ is the dimensionless Fermi momentum. The
lepton chemical potential is:
\begin{equation}
    \mu_l = \sqrt{k_{F,l}^2 + m_l^2}, \quad l = e, \mu.
\end{equation}
The baryon chemical potential is obtained from the energy density as
\begin{equation}
    \mu_B = \frac{\partial \varepsilon_B}{\partial n_B}.
\end{equation}
The total energy density and pressure of the nucleonic core are calculated as
\begin{equation}
    \varepsilon_\mathrm{tot} = \varepsilon_B + \varepsilon_l, \quad
    P_\mathrm{tot} = P_B + P_l.
    \label{leptonEoS}
\end{equation}

\subsubsection{Hyperonic core}

When hyperons are included, the beta-equilibrium conditions are
generalized as
\begin{equation}
    \mu_B = b_B \mu_n - q_B \mu_e,
\end{equation}
where $b_B$ and $q_B$ denote the baryon number and electric charge of
baryon $B$, respectively. Explicitly, for the octet baryons:
\begin{subequations}
\begin{align}
    \mu_\Lambda   &= \mu_n, \\
    \mu_{\Sigma^-}&= \mu_n + \mu_e, \\
    \mu_{\Sigma^0}&= \mu_n, \\
    \mu_{\Sigma^+}&= \mu_n - \mu_e, \\
    \mu_{\Xi^-}   &= \mu_n + \mu_e, \\
    \mu_{\Xi^0}   &= \mu_n.
\end{align}
\end{subequations}
A hyperon species appears in matter when its chemical potential
satisfies the above equilibrium condition at a given baryon density.
The charge neutrality condition is modified to account for the charged
hyperons:
\begin{equation}
    n_p + n_{\Sigma^+} = n_e + n_\mu + n_{\Sigma^-} + n_{\Xi^-}.
    \label{eq:chargeneutrality_H}
\end{equation}
The total energy density and pressure have the same form as in the
nucleonic case Eq.~\eqref{leptonEoS}, where $\varepsilon_B$ and $P_B$ now
include contributions from all baryon species. The self-consistent
solution of the meson field equations, beta-equilibrium conditions, and
charge neutrality determines the composition and EoS of the hyperonic
core as a function of baryon density.

%--------------------------------------------------------------------
\begin{figure}[t]
    \centering
    \includegraphics[width=\columnwidth]{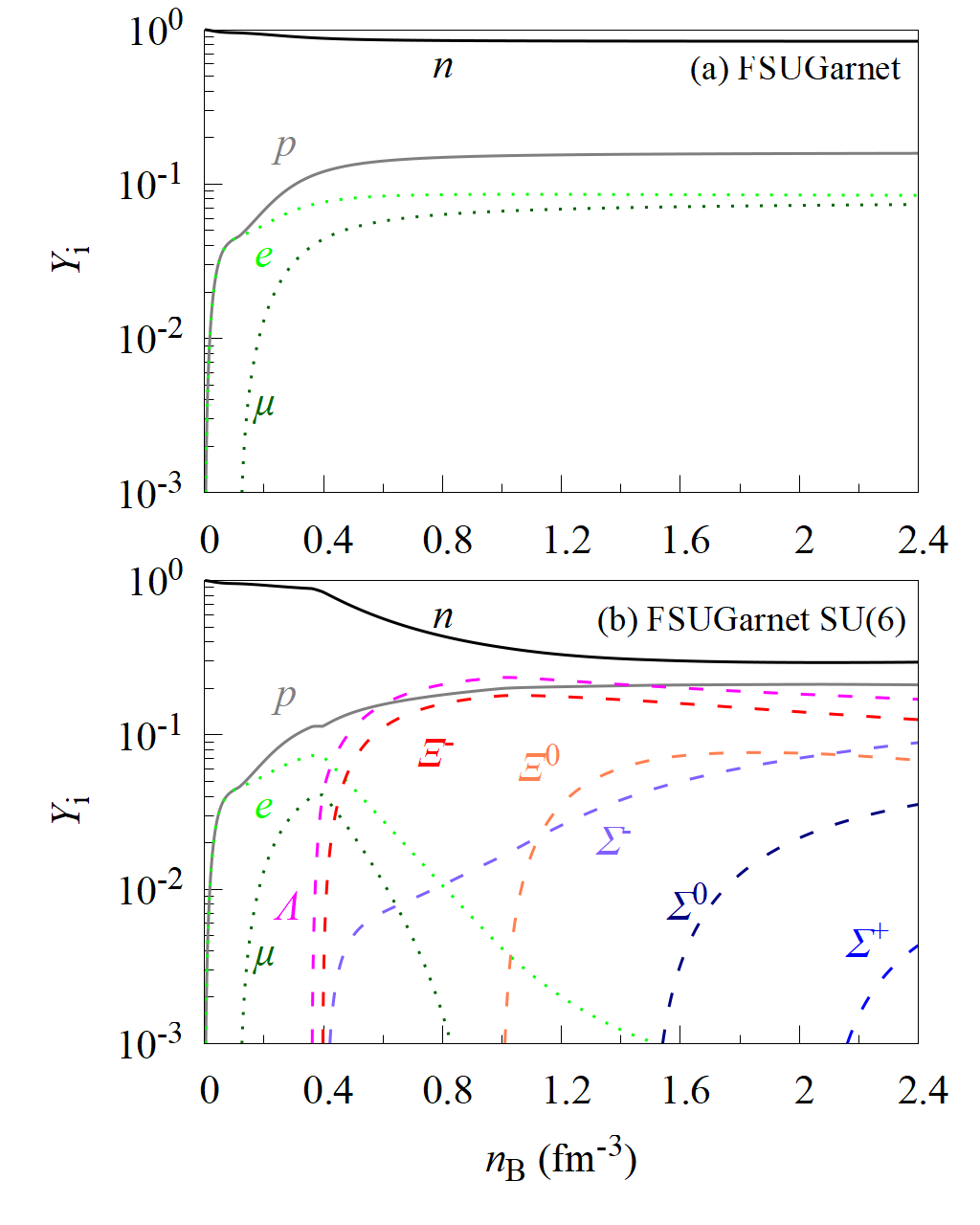}\vspace{-5mm}
    \caption{Particle fractions as a function of baryon number 
    density for (a) the nucleonic EoS and (b) the hyperonic EoS based on FSUGarnet parameter set.}
    \label{fig:particle_fraction}
\end{figure}
%--------------------------------------------------------------------

\begin{figure}[t]
    \centering
    \includegraphics[width=\columnwidth]{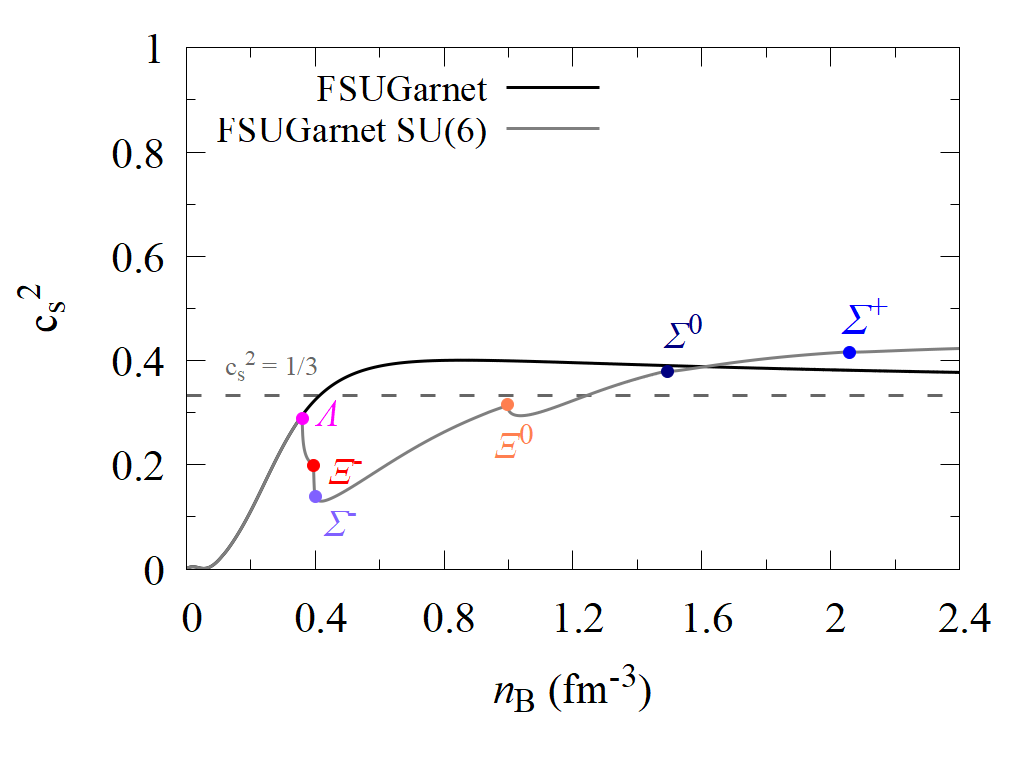}\vspace{-5mm}
    \caption{Speed of sound squared $c_s^2$ as a function of baryon
    number density $n_B$ for FSUGarnet (black) and FSUGarnet+SU(6)
    (gray). The dashed horizontal line indicates the conformal limit
    $c_s^2 = 1/3$. In the FSUGarnet+SU(6) case, the sequential onset
    of $\Lambda$ (at $n_B \sim 0.36\,\mathrm{fm}^{-3}$), $\Xi^-$
    (at $n_B \sim 0.395\,\mathrm{fm}^{-3}$), $\Sigma^-$
    (at $n_B \sim 0.40\, \mathrm{fm}^{-3}$), $\Xi^0$
    (at $n_B \sim 0.995\,\mathrm{fm}^{-3}$), $\Sigma^0$
    (at $n_B \sim 1.492\,\mathrm{fm}^{-3}$), and $\Sigma^+$
    (at $n_B \sim 2.058\,\mathrm{fm}^{-3}$) are marked, each
    corresponding to a visible softening of the EoS.}
    \label{fig:cs2}
\end{figure}

The resulting particle fractions as a function of baryon density are shown in Fig.~\ref{fig:particle_fraction} for the nucleonic (upper panel) and hyperonic (lower panel) cases. In the nucleonic EoS, only protons, neutrons, electrons, and muons are present. In the hyperonic EoS, hyperons appear sequentially with increasing density, as shown in Fig.~\ref{fig:particle_fraction}. The detailed onset densities for each species are indicated in the caption of Fig.~\ref{fig:cs2}. Note that the effective mass in the FSUGarnet parameter set becomes zero or negative at extremely high density, leading to unphysical states. Therefore, we performed calculations only up to $2.5$ $\text{fm}^{-3}$. We also examine the speed of sound $c_s^2$ for the two EoSs considered in this work. The speed of sound is defined as
\begin{equation}
    c_s^2 = \frac{\partial P}{\partial \varepsilon}.
\end{equation}
A value of $c_s^2 = 1/3$ corresponds to the conformal limit expected
to be approached asymptotically at sufficiently high
densities. We verify that $c_s^2 \leq 1$ throughout
the density range considered, ensuring that causality is satisfied for both
EoSs. Figure~\ref{fig:cs2} shows $c_s^2$ as a function of baryon
number density $n_B$ for both FSUGarnet (thick black line)
and FSUGarnet+SU(6) (thin gray line).

%--------------------------------------------------------------------
%--------------------------------------------------------------------

From Fig.~\ref{fig:cs2}, we find that, for the pure nucleonic FSUGarnet model, $c_s^2$ increases monotonically and exceeds the conformal limit $c_s^2 = 1/3$ beyond $n_B \sim 0.36\,\mathrm{fm}^{-3}$, indicating a stiff EoS with no phase transition. In the FSUGarnet+SU(6) model, the sequential onset of hyperons introduces significant softenings. The first drop near $n_B
\sim 0.36\,\mathrm{fm}^{-3}$ corresponds to the appearance of $\Lambda$, followed closely by $\Xi^-$ and $\Sigma^-$, which open new degrees of freedom and reduce the Fermi pressure. A second softening occurs near $n_B \sim 0.995\,\mathrm{fm}^{-3}$ with the onset of $\Xi^0$, followed by $\Sigma^0$ and $\Sigma^+$ at higher densities.
 
The complete EoS, obtained by matching the crust and core
contributions, serves as input to the stellar structure equations described in the following subsection.

\subsection{Modeling of neutron star structure}

\subsubsection{Static neutron star}

Neutron stars are compact objects with extremely strong gravitational fields, for which general relativistic effects cannot be neglected. The structure of such relativistic compact objects is determined by solving the Einstein field equations~\cite{Einstein1916}. Throughout this section, we adopt units in which $G=c=1$. For non-rotating (spherically symmetric) configurations, which also serve as initial conditions for rotating solutions, the structure is governed by the Tolman-Oppenheimer-Volkoff (TOV) equation~\cite{Oppenheimer1939}:
\begin{subequations}
    \begin{align}
        \frac{\dd P}{\dd r} &= - (\varepsilon+P)\frac{M+4\pi r^3P}{r(r-2M)},\\
        \frac{\dd M}{\dd r} &= 4\pi r^2 \varepsilon.
    \end{align}
\end{subequations}
These equations are integrated numerically from the center outward until the pressure vanishes at the stellar surface.

%--------------------------------------------------------------------
\begin{figure}[t]
    \centering
    \includegraphics[width=\columnwidth]{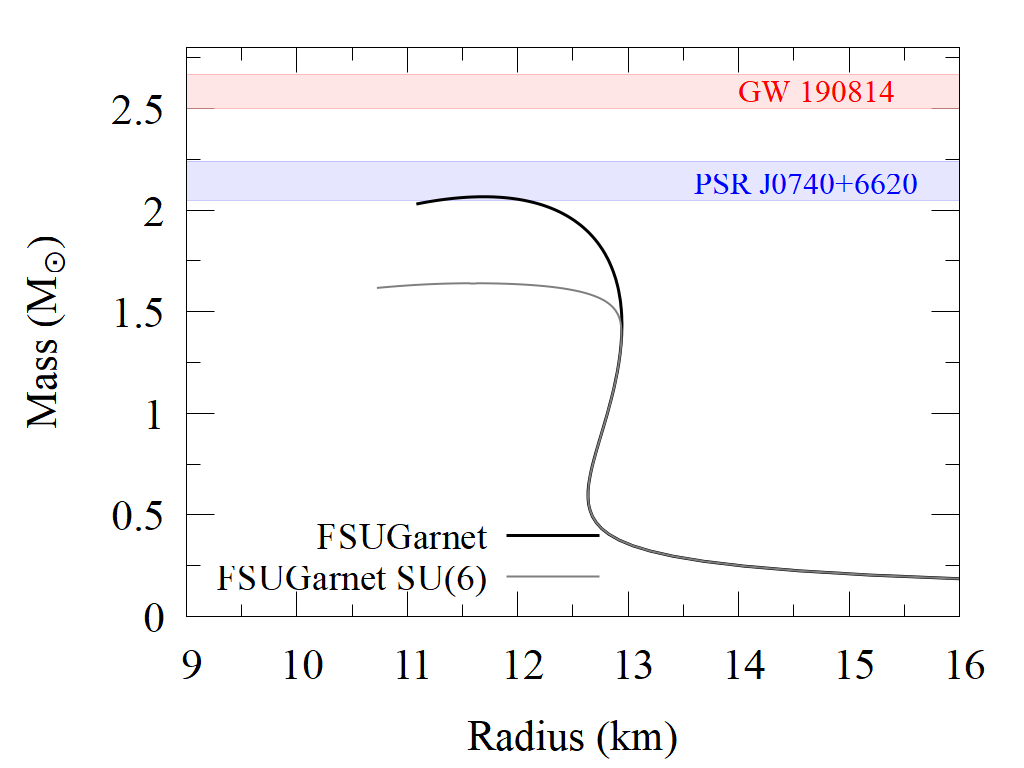}\vspace{-2mm}
    \caption{Mass-radius relations for static neutron stars computed with the FSUGarnet and FSUGarnet+SU(6) equations of state. Shaded bands denote observational constraints from PSR~J0740+6620~\cite{Cromartie2020} and GW\,190814~\cite{Abbott2020}.}
    \label{fig:tov}
\end{figure}
%--------------------------------------------------------------------

The $M$-$R$ curves obtained using FSUGarnet and FSUGarnet+SU(6) are shown in Fig.~\ref{fig:tov}. When only nucleonic degrees of freedom are considered, the maximum mass satisfies the observational constraint from PSR~J0740+6620. However, upon inclusion of hyperonic degrees of freedom, the maximum mass is significantly reduced, and the equation of state can no longer account for the observed mass.

\subsubsection{Rotating neutron star}

To model rotating neutron stars, we employ the
KEH method~\cite{Komatsu1989a} to construct
rotating neutron star configurations.
The metric for a stationary, axisymmetric spacetime is written as
\begin{equation}
    \begin{aligned}
    \dd s^2 &= -e^{\gamma+\varrho}\dd t^2 + e^{2\alpha}(\dd r^2 + r^2\dd\theta^2) \\
    &\quad + e^{\gamma-\varrho}r^2 \sin^2\theta(\dd\phi-\omega\dd t)^2,
    \end{aligned}
    \label{eq:metric}
\end{equation}
where $\gamma(r,\theta)$, $\varrho(r,\theta)$, and $\alpha(r,\theta)$ are metric potentials, and $\omega(r,\theta)$ is the frame-dragging angular velocity. The KEH method reformulates the Einstein field equations into a coupled set of elliptic partial differential equations and
one ordinary differential equation:
\begin{subequations}
    \begin{align}
        &\nabla^2[\varrho e^{\gamma/2}] = S_{\varrho}(r,\mu), \\
        &\left( \nabla^2 + \frac{1}{r} \partial_r - \frac{\mu}{r^2} \partial_{\mu} \right) [\gamma e^{\gamma/2}] = S_{\gamma}(r,\mu), \\
        &\left( \nabla^2 + \frac{2}{r} \partial_r - \frac{2\mu}{r^2} \partial_{\mu} \right) [\omega e^{(\gamma-2\varrho)/2}] = S_{\omega}(r,\mu), \\
        & \frac{\partial \alpha}{\partial \mu} = S_\alpha(r,\mu),
    \end{align}
    \label{eq:KEHES}
\end{subequations}
where $\mu \equiv \cos\theta$, and $S_\varrho$, $S_\gamma$,
$S_\omega$, and $S_\alpha$ are source terms determined by the
matter distribution and the metric potentials.

In this work, the differential rotation law is defined as
\begin{equation}
    j(\Omega) \equiv u^t u_\phi = A^2(\Omega_c - \Omega),
    \label{eq:rotlaw}
\end{equation}
where $A$ denotes the differential rotation length scale, and $\Omega_c$ is the 
angular velocity at the stellar center. In the limit $A \to \infty$ the rotation 
approaches uniform rotation, whereas smaller values of $A$ correspond to stronger 
differential rotation.

We adopt the CST approach~\cite{Cook1992}, which maps radial infinity to the finite coordinate $s = 1$ via
\begin{equation}
    r = r_e \Bigl( \frac{s}{1-s} \Bigr),
    \label{eq:CSTmap}
\end{equation}
where $r_e$ is the equatorial radius. Under this compactification, angular velocities are rescaled to define a new computational space:
\begin{subequations}
    \begin{align}
        &\omega = \hat{\omega} /r_e, \\
        &\Omega = \hat{\Omega}/r_e.
    \end{align}    
    \label{eq:rescale1}
\end{subequations}
The rescaled source terms are provided in the original CST papers~\cite{Cook1994b,Cook1994}.
To evaluate the hydrostatic equilibrium, we introduce the rescaled metric potentials $\hat{\alpha}$, $\hat{\varrho}$, $\hat{\gamma}$, and $\hat{A}$:
\begin{subequations}
    \begin{align}
        &\alpha = r^2_e \hat{\alpha}, \\
        &\varrho = r^2_e \hat{\varrho}, \\
        &\gamma = r^2_e \hat{\gamma}, \\
        &A = r_e \hat{A}.
    \end{align}    
    \label{eq:rescale2}
\end{subequations}
Note that the source terms $S_\varrho$, $S_\gamma$, $S_{\hat\omega}$, and $S_\alpha$ in the original CST formulation are expressed in a mixed notation: $\hat{\omega}$ and $\hat{\Omega}$ appear in their rescaled forms, while the remaining metric potentials $\varrho$, $\gamma$, and $\alpha$ retain their unscaled definitions. Careful attention to this convention is required during numerical implementation.

The equation of hydrostatic equilibrium follows from the
conservation of energy-momentum, $\nabla_aT^{ab} = 0$, and can be written in an integrated form as follows:
\begin{equation}
    \ln{H} + \frac{\hat{\varrho}+\hat{\gamma}}{2}r^2_e + \frac{1}{2}\ln{(1-v^2)} - \frac{1}{2}\hat{A}^2(\hat{\Omega}-\hat{\Omega}_c)^2 = C,
    \label{eq:hydro}
\end{equation}
where $C$ is a constant. The specific enthalpy $H$ is defined as
\begin{equation}
    \ln H = \int \frac{\dd P}{\varepsilon + P},
    \label{eq:enthalpy}
\end{equation}
and the proper velocity $v$ of the fluid as measured by a zero-angular-momentum observer 
is~\cite{Butterworth1976}
\begin{equation}
    v = (\hat{\Omega} - \hat{\omega})
    \Bigl( \frac{s}{1-s} \Bigr)\sin\theta\,e^{-r_e^2\hat{\varrho}}.
    \label{eq:velocity}
\end{equation}
The specific enthalpy $H$ is determined from the EoS. Solving Eq.~\eqref{eq:hydro} for $\hat{\Omega}$ together with the differential rotation law Eq.~\eqref{eq:rotlaw} yields:
\begin{equation}
    \hat{A}^2(\hat{\Omega}_c - \hat{\Omega})
    = \frac{(\hat{\Omega} -  \hat{\omega})\,s^2 \sin^2 \theta \,e^{-2 r^2_e \hat{\varrho}}}{(1-s)^2 - (\hat{\Omega} - \hat{\omega})^2\,s^2 \sin^2 \theta \,e^{-2 r^2_e \hat{\varrho}}}.
    \label{eq:rot}
\end{equation}

The boundary conditions are imposed at the stellar surface, specifically at the pole and the equator.
At the pole, the fluid lies on the rotation axis, so $j = 0$ by symmetry; from the rotation law~\eqref{eq:rotlaw} this implies
$\Omega = \Omega_c$.
Moreover, the energy density and proper velocity vanish there,
$\ln H = v = 0$, so that Eq.~\eqref{eq:hydro} reduces to
\begin{equation}
    \frac{\hat{\varrho}_p + \hat{\gamma}_p}{2}\,r_e^2 = C,
    \label{eq:pole}
\end{equation}
where the subscript $p$ denotes evaluation at the pole. At the center, the proper velocity vanishes ($v_c = 0$) and 
the differential rotation term contributes nothing 
($\hat{\Omega} = \hat{\Omega}_c$), so that 
Eq.~\eqref{eq:hydro} is reduced to
\begin{equation}
    \ln H_c + \frac{\hat{\varrho}_c + \hat{\gamma}_c}{2}\,r_e^2 = C,
    \label{eq:center}
\end{equation}
where the subscript $c$ denotes evaluation at the center, and $H_c$ is the specific enthalpy at the center, 
determined from the prescribed central energy density 
$\varepsilon_c$ via the EoS. Subtracting 
Eq.~\eqref{eq:center} from Eq.~\eqref{eq:pole} yields
\begin{equation}
    r_e^2 = \frac{2\ln H_c}
    {\hat{\varrho}_p + \hat{\gamma}_p 
     - \hat{\varrho}_c - \hat{\gamma}_c},
    \label{eq:re}
\end{equation}
which is updated at each iteration of the self-consistent field loop until convergence is achieved. The equatorial radius $r_e$ is determined from the equilibrium condition at the center. 
 
The hydrostatic equilibrium condition and the rotation condition at the equator (where $\sin\theta_e = 1$) give
\begin{subequations}
    \begin{align}
        &\frac{\hat{\varrho}_e + \hat{\gamma}_e}{2}\,r_e^2
        + \frac{1}{2}\ln(1 - v_e^2)
        - \frac{1}{2}\hat{A}^2(\hat{\Omega}_e - \hat{\Omega}_c)^2 = C, \label{eq:equator_hydro} \\
        &\hat{A}^2(\hat{\Omega}_c - \hat{\Omega}_e)
        = \frac{(\hat{\Omega}_e -  \hat{\omega}_e)\,s_e^2  \,e^{-2 r^2_e \hat{\varrho}_e}}{(1-s_e)^2 - (\hat{\Omega}_e - \hat{\omega}_e)^2\,s_e^2  \,e^{-2 r^2_e \hat{\varrho}_e}}, \label{eq:equator_rot} 
    \end{align}
\end{subequations}
where the subscript $e$ denotes evaluation at the equator.
Subtracting Eq.~\eqref{eq:pole} from Eq.~\eqref{eq:equator_hydro} yields
\begin{equation}
    \bigl(\hat{\varrho}_p + \hat{\gamma}_p
          - \hat{\varrho}_e - \hat{\gamma}_e\bigr)r_e^2
    = \ln(1 - v_e^2)
    - \hat{A}^2(\hat{\Omega}_e - \hat{\Omega}_c)^2.
    \label{eq:bc_combined}
\end{equation}
By solving Eqs.~\eqref{eq:equator_rot} and~\eqref{eq:bc_combined} numerically, 
we obtain $\hat{\Omega}_e$ and $\hat{\Omega}_c$. Using Eq.~\eqref{eq:rot}, 
we can then determine the angular velocity $\hat{\Omega}(s,\mu)$ at arbitrary points. 

We can also derive the enthalpy at each point by subtracting Eq.~\eqref{eq:hydro} 
from Eq.~\eqref{eq:pole}:
\begin{equation}
    \begin{aligned}
        \ln{H} &= \frac{1}{2}\bigl(\hat{\varrho}_p + \hat{\gamma}_p
          - \hat{\varrho} - \hat{\gamma}\bigr)r_e^2 \\
          &\quad - \frac{1}{2}\ln(1-v^2)+\frac{1}{2}\hat{A}^2(\hat{\Omega}-\hat{\Omega}_c)^2.
    \end{aligned}
\end{equation}
The hydrostatic equilibrium condition thus yields the enthalpy $H$ at each point, from which the pressure $P$ and energy density $\varepsilon$ are computed via 
the EoS through Eq.~\eqref{eq:enthalpy}. 

Once a self-consistent solution is obtained, the physical metric potentials and angular velocities are recovered by inverting Eqs.~\eqref{eq:rescale1} and \eqref{eq:rescale2}. We then compute several global quantities characterizing the rotating neutron star. All volume integrals are evaluated using Simpson's rule over the compactified grid. The circumferential radius at the equator is defined by
\begin{equation}
    R_\text{e}=r_e e^{(\gamma_e + \varrho_e)/2}.
\end{equation}
The gravitational mass $M$, proper mass $M_p$, angular momentum $J$, and rotational kinetic energy $T$ are given by
\begin{widetext}
 \begin{subequations}
     \begin{align}
             &M =\int (-2T^t_t + T^\mu_\mu)\sqrt{-g} \dd^3x = 4\pi r_e^3 \int_0^1 \frac{s^2 \dd s}{(1-s)^4}
    \int_0^1 \dd \mu e^{2\alpha+\gamma}
    \left[
        \frac{\varepsilon + P}{1-v^2}
        \left(1 + v^2 + \frac{2sv \omega r_e}{1-s} e^{-\varrho} \sqrt{1-\mu^2} \right)
        + 2P
    \right], \label{eq:mass} \\
    &    M_p = \int \rho_0 u^t \sqrt{-g} \dd^3x = 4\pi r_e^3 \int_0^1 \frac{s^2 \dd s}{(1-s)^4}
    \int_0^1 \dd \mu 
    e^{2\alpha+(\gamma-\varrho)/2}\,
    \frac{\varepsilon}{\sqrt{1-v^2}}\,,
    \label{eq:massp} \\
        &J = \int T^t_\phi \sqrt{-g} \dd^3 x = 4\pi r_e^4 \int_0^1 \frac{s^3 \dd s}{(1-s)^5} \int_0^1 \dd\mu
    \sqrt{1-\mu^2}\,
    e^{2\alpha+\gamma-\varrho}\,
    \frac{(\varepsilon+P)\,v}{1-v^2}\,,
    \label{eq:angmom} \\
        &T = \frac{1}{2}
\int \Omega T^t_\phi \sqrt{-g} \dd^3 x = 2\pi r_e^4 \int_0^1 \frac{s^3 \dd s}{(1-s)^5} \int_0^1\dd\mu
    \sqrt{1-\mu^2} \, e^{2\alpha+\gamma-\varrho}\,
    \frac{(\varepsilon+P)\,v \Omega }{1-v^2} \,,
    \label{eq:Trot}
     \end{align}
 \end{subequations}
\end{widetext}

For the metric ansatz of Eq.~\eqref{eq:metric}, the determinant is
$\sqrt{-g} = e^{2\alpha+\gamma}\, r^2 \sin\theta$, so that
$\sqrt{-g}\,\dd^3x = e^{2\alpha+\gamma}\, r^2\,\dd r\,\dd\mu\,\dd\phi$
with $\mu=\cos\theta$. Under the compactified radial coordinate
$r = r_e\, s/(1-s)$ and after performing the trivial $\phi$ integration,
this reduces to the explicit forms. The gravitational binding energy is then defined as
\begin{equation}
    |W| = M_p - M + T,
    \label{eq:Wbind}
\end{equation}
and the ratio,
\begin{equation}
    \beta \equiv \frac{T}{|W|},
    \label{eq:ToverW}
\end{equation}
serves as a key diagnostic for rotational instabilities: 
the Newtonian secular and dynamical instability thresholds occur at $\beta \simeq 0.14$ and $\beta \simeq 0.27$, respectively~\cite{Chandrasekhar1969,Houser1994}. These thresholds are derived for incompressible Newtonian bodies and serve as approximate reference values; the exact thresholds for relativistic, compressible stars may differ.

To compute the Kepler frequencies, the solution of the geodesic equation yields two possible values for the orbital three-velocity~\cite{Bardeen1970,Cook1994b}:
\begin{widetext}
   \begin{equation}
        \tilde{v}_e =
        \frac{e^{-\varrho_e}  r_e s^2_e (\partial_s \omega_e) \pm
        \sqrt{e^{-2\varrho_e} r^2_e s^4_e (\partial_s \omega_e)^2
        + 2s_e(1-s_e) \{(\partial_s \gamma_e) + (\partial_s \varrho_e)\} + s^2_e(1-s_e)^2 \{(\partial_s \gamma_e) - (\partial_s \varrho_e)\}^2}
}{2 + s_e(1-s_e)\{(\partial_s \gamma_e) - (\partial_s \varrho_e)\} }.
\end{equation} 
\end{widetext}
The positive sign corresponds to corotating orbits, while the negative sign corresponds to counter-rotating orbits with respect to a zero-angular-momentum observer. The Keplerian angular frequencies at the equator derived in the general relativistic formalism are then defined by~\cite{Morsink1999}:
\begin{equation}
    f_{Ke} = \frac{1}{2 \pi} \left( \tilde{v}_e \frac{e^{\varrho_e}}{r_e} + \omega_e \right).
    \label{Eq:Keplar_GR}
\end{equation}

\section{Computational details}
\label{Sec:Details}

A critical numerical consideration in the KEH method is the choice of grid resolution. In our previous work on uniformly rotating 
stars~\cite{Kwon2026,Kwon2026b}, the grid size had negligible effect on the computed quantities of interest. In the present study, however, the degree of differential rotation, parametrized 
by $\hat{A}^{-1}$, significantly affects the internal structure 
of the star, and an insufficient number of grid points fails 
to resolve the strongly deformed density distribution. 
Furthermore, extremely oblate configurations arising from 
strong differential rotation require a finer mesh in the $\mu = \cos\theta$ direction.

%--------------------------------------------------------------------------
\begin{table}[h]
\caption{Results of the KEH calculations for $r_\text{ratio}=r_\text{p}/r_\text{e}=0.2$ and central density $\rho_c = 1.0 \times 10^{15}$\,g/cm$^3$ ($n_c\simeq3.35n_0$) with FSUGarnet+SU(6) EoS, with five mesh resolutions, $65\times65$, $129 \times 129$, $257 \times 257$, $513\times513$, and $1025\times1025$.
}\vspace{2mm}
\begin{tabular}{lccc}
\hline\hline
\# of grid points & Mass ($M_\odot$) & $R_e$ (km) & $f_e$ (Hz) \\
\hline
 $1025\times1025$  & 2.46598 & 12.112804 & 1565.5240 \\
 $513\times513$  & 2.46592 & 12.112529 & 1565.5575 \\
 $257\times257$  & 2.46564 & 12.111281 & 1565.7119 \\
 $129\times129$  & 2.46453 & 12.106264 & 1566.3409 \\
 $65\times65$  & 2.46087 & 12.090040 & 1568.3732 \\ 
\hline\hline
\end{tabular}\label{tab:resolutions}
\end{table}
%--------------------------------------------------------------------------

Table~\ref{tab:resolutions} shows the sensitivity of the computed global quantities to grid resolution for a representative model with central density 
$\rho_c = 1.0\times10^{15}$\,g\,cm$^{-3}$, $\hat{A}^{-1}=1$, and 
axis ratio $r_{\text{ratio}} = r_p/r_e = 0.2$. 
The $129\times129$ grid shows minor deviations from finer grids, but the differences remain well within acceptable numerical tolerance. 

To quantify this convergence behavior, we evaluate the observed 
convergence rate using the equatorial radius $R_e$ and rotational 
frequency $f_e$ as
\begin{equation}
    p_n = \log_2\!\left(\frac{|Q_{n}-Q_{n-1}|}{|Q_{n+1}-Q_{n}|}\right),
\end{equation}
where $Q_n$ denotes the value of a global quantity ($R_e$ or $f_e$) 
computed on the $n$-th mesh, with $n$ increasing from the coarsest 
($65\times65$) to the finest ($1025\times1025$) resolution.
The convergence factor $p_n$ for $R_e$ and $f_e$ remains close to 2, 
indicating that the numerical results are consistent with the designed 
second-order accuracy.

We therefore adopt a $129 \times 129$ grid for large-scale surveys of neutron star sequences
(e.g., mass-radius relations), where computational efficiency is prioritized, and a $513 \times 513$ grid for detailed analyses of individual stellar structures and density profiles. Legendre polynomials up to $P_{32}(\mu)$ are used in this study.

As a code validation, our results for uniformly rotating configurations are compared against those of 
Stergioulas \& Friedman~\cite{Stergioulas1995} and show 
good agreement, while results for differentially rotating configurations are found to be in good agreement with those of Morrison \textit{et al.}~\cite{Morrison2004}.

For uniformly rotating stars, initializing the 
self-consistent field (SCF) iteration with the 
corresponding TOV solution is sufficient for convergence. 
For differentially rotating configurations, however, 
models with small axis ratios $r_{\text{ratio}}$ and large 
$\hat{A}^{-1}$ are highly deformed, and the SCF iteration 
may fail to converge when initialized from a spherical 
solution. To address this, we employ a 
\textit{continuation method}: starting from a 
converged solution at a moderate axis ratio, we 
decrease $r_{\text{ratio}}$ in steps of $0.001$, using 
the metric potentials of the previous converged 
solution as the initial guess for the next iteration. 
This approach ensures stable convergence across the 
full range of deformation parameters explored in 
this work.

For the SU(6)-based hyperonic equation of state 
at high densities, convergence of the meson mean-field 
equations is similarly improved by initializing the 
mean-field values from the solution at the preceding 
density step, rather than from zero. This strategy ensures stable convergence of the hadronic field equations throughout the density range of interest.

%--------------------------------------------------------------------
\begin{figure}[t]
    \centering
    \includegraphics[width=\columnwidth]{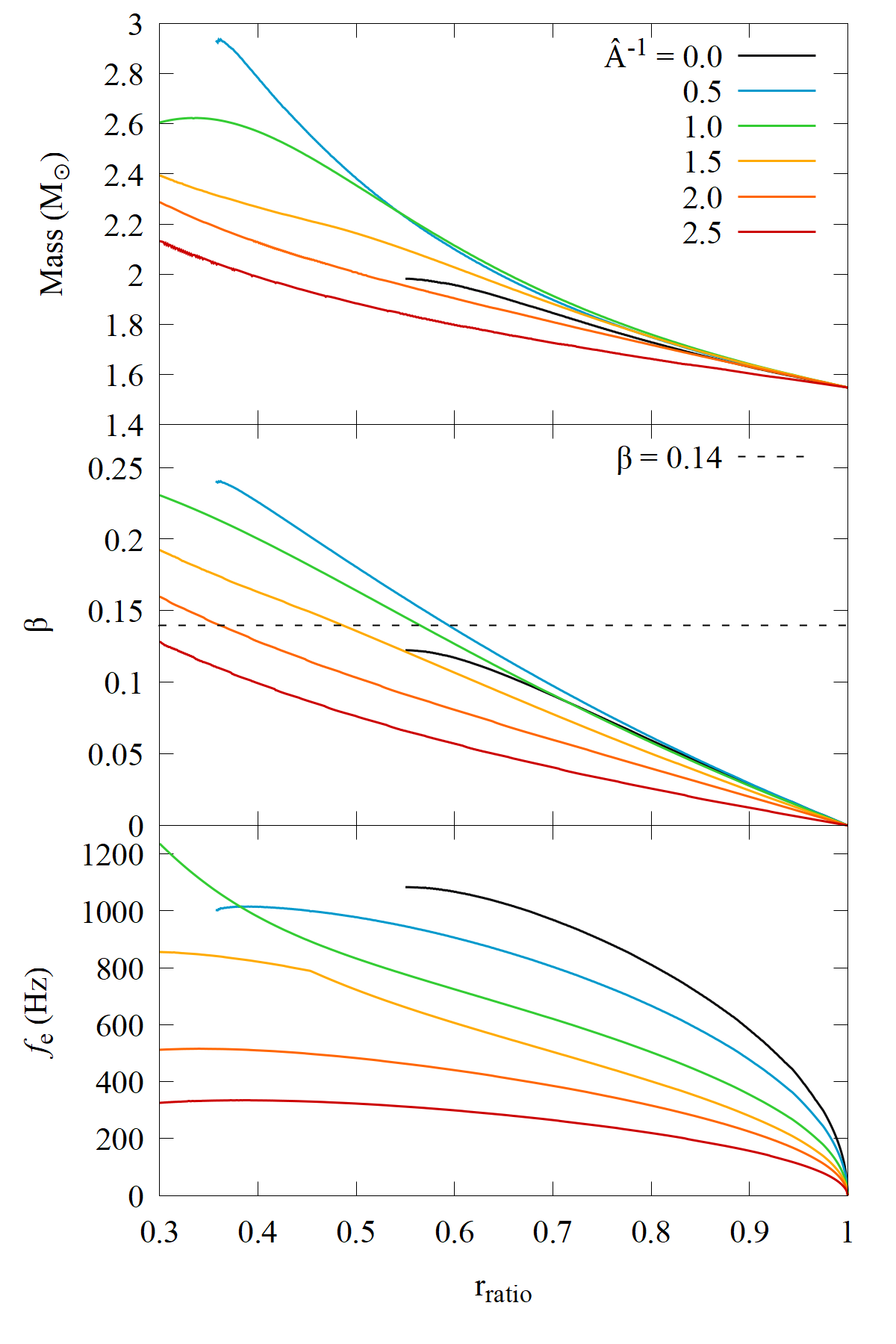}\vspace{-5mm}
    \caption{Properties of differentially rotating neutron stars as a function 
of the axis ratio $r_\text{ratio}$ for various differential rotation 
parameters $\hat{A}^{-1} = 0.0$ - $2.5$. 
Top: gravitational mass $M$ in units of $M_\odot$. 
Middle: ratio of rotational kinetic energy to gravitational potential energy 
$\beta$. Bottom: rotational frequency at the equator $f_e$ in Hz.}
    \label{fig:properties}
\end{figure}
%--------------------------------------------------------------------

\section{Results and Discussion}\label{Sec:Results}

\subsection{Properties of differentially rotating neutron stars}

First, we present general properties of differentially rotating neutron star configurations. Figure~\ref{fig:properties} shows the gravitational mass $M$, the ratio of rotational kinetic energy to gravitational potential energy $\beta$, and the spin frequency at the equator $f_e=\Omega_e / 2\pi$ as functions of $r_{\text{ratio}}$ at a central density of $\rho_c = 1.0 \times 10^{15}$ $\text{g}/\text{cm}^3$, computed using the FSUGarnet+SU(6) EoS. Each sequence is computed up to the point where the equatorial spin frequency $f_e$ reaches the Kepler frequency $f_\text{Ke}$, beyond which mass shedding occurs.

A large value of $\hat{A}^{-1}$ corresponds to strong differential rotation, and one might naively expect this to produce a higher equatorial spin frequency. However, strong differential rotation concentrates rapid rotation in the core region, so that the surface angular velocity is conversely reduced. In fact, uniform rotation $(\hat{A}^{-1}=0)$ yields the highest equatorial spin frequency among stable, oblate neutron star configurations $(\beta<0.14)$.

For moderate differential rotation, $\hat{A}^{-1} \approx 1$, $\beta$ can exceed 0.14, allowing the neutron star to become significantly deformed and its internal structure to be substantially altered; as a consequence, a higher spin frequency at the equator is achieved. This indicates that neither the weakest nor the strongest degree of differential rotation produces the fastest equatorial surface rotation. Rather, it is an intermediate degree of differential rotation that enables the equilibrium configuration with the highest rotation rate.

Regarding the gravitational mass, among the values of $\hat{A}^{-1}$ considered here, $\hat{A}^{-1}=0.5$ yields the largest maximum mass. For large values of $r_{\text{ratio}}$, the supported mass tends to decrease as $\hat{A}^{-1}$ departs from 0.5 in either direction. Furthermore, for strongly deformed configurations, nearly uniform rotators are found to exceed the Kepler frequency at the surface and thus 
%cannot form stable 
do not admit equilibrium configurations, whereas stars with strong differential rotation remain below the Keplerian limit even at high deformation.

\subsection{Maximum Mass of Neutron Stars}

\subsubsection{PSR\,J0740$+$6620}

\begin{figure}[t]
    \centering
    \includegraphics[width=\columnwidth]{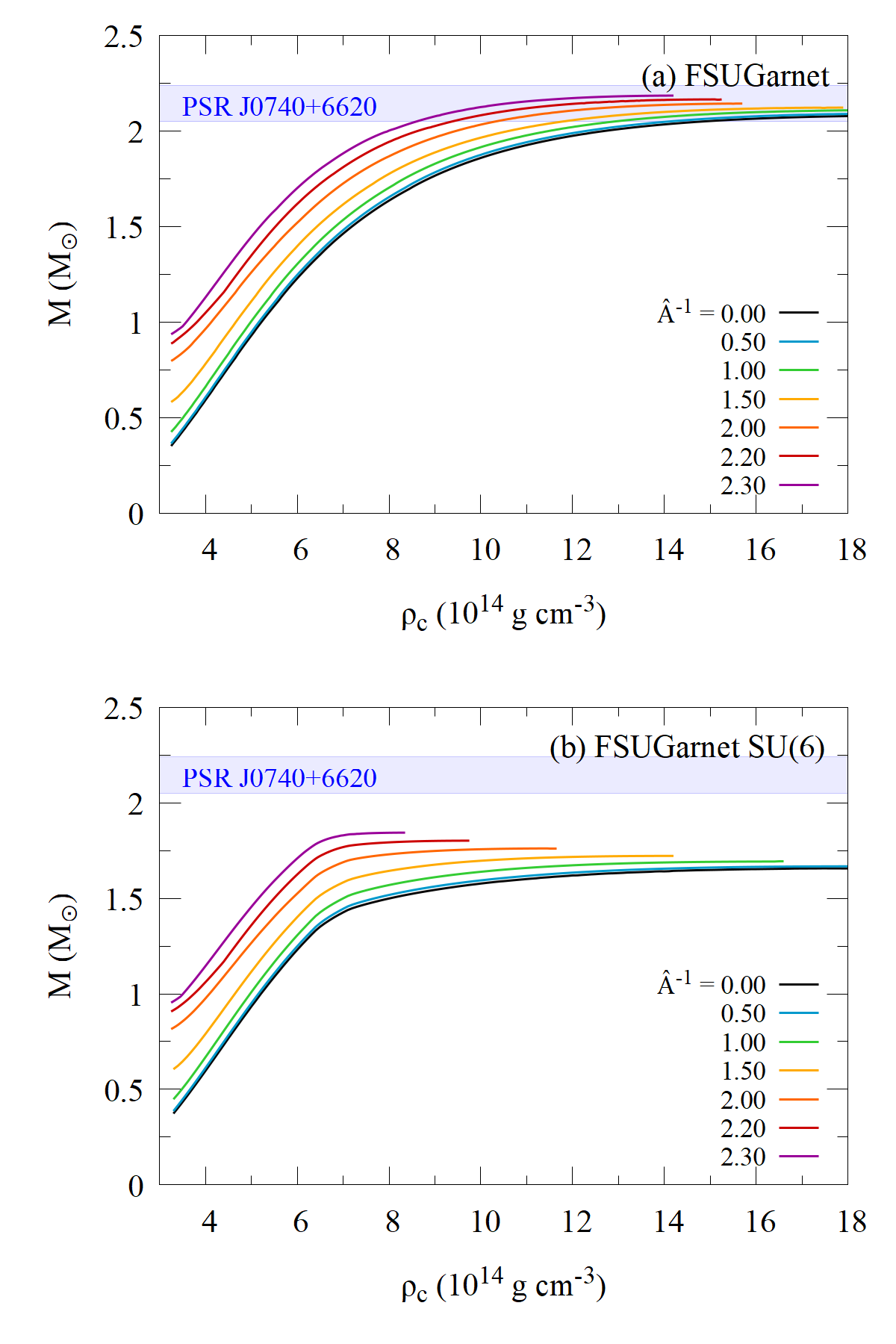}\vspace{-5mm}
    \caption{Mass-central density relations for differentially rotating neutron stars at a fixed spin frequency of $346 \, \text{Hz}$, computed with the FSUGarnet and FSUGarnet+SU(6) EoS. Each curve corresponds to a different degree of differential rotation, parametrized by $\hat{A}^{-1}=0.00,0.50,1.00,1.50,2.00,2.20,$ and $2.30$, where larger values indicate stronger differential rotation. The horizontal shaded band indicates the mass range of PSR~J0740$+$6620.}
    \label{fig:PSR}
\end{figure}

Next, we investigate whether equilibrium configurations satisfying 
the observational constraints of PSR~J0740+6620 can be constructed 
using the FSUGarnet$+$SU(6) hyperonic EoS. PSR~J0740+6620, rotating at 346\,Hz, 
serves as one of the most stringent constraints on the neutron star 
maximum mass and lies at the heart of the hyperon puzzle. 
The mass-central density relations shown in Fig.~\ref{fig:PSR} are computed for varying  $\hat{A}^{-1}$, with central densities ranging from $\rho_c = 3.3\times10^{14}$\,g\,cm$^{-3}$ up to the turning-point
configuration defined by $\partial M / \partial \rho_c = 0$. Although the turning-point criterion ($\partial M / \partial \rho_c = 0$ at fixed angular momentum) is rigorously established only for static and uniformly rotating stars~\cite{Friedman1988}, and even there it provides only a sufficient---not necessary---condition for secular axisymmetric instability~\cite{Takami2011}, we adopt it here as an approximate stability limit, since a rigorous criterion for differentially rotating configurations
is not readily available. 
Dynamical evolutions indicate that the true stability boundary lies slightly on the low-density side of the turning point but remains close to it, so that the turning point still provides a
reasonable estimate of the stability limit for differentially rotating neutron stars~\cite{Weih2018}.
%--------------------------------------------------------------------
%--------------------------------------------------------------------

At a fixed spin frequency at the equator of 346\,Hz, we find that larger values 
of $\hat{A}^{-1}$ yield higher maximum masses. However, when 
$\hat{A}^{-1}$ becomes too large, equilibrium configurations with 
an equatorial spin frequency of $f_e = 346$\,Hz can no longer be 
constructed. This is illustrated by the $\hat{A}^{-1} = 2.5$ case 
in Fig.~\ref{fig:properties}: a large degree of differential rotation implies that the central angular velocity greatly exceeds the surface angular velocity, making it impossible to achieve equilibrium configurations with sufficiently large surface rotation rates.

For the FSUGarnet EoS (Fig.~\ref{fig:PSR}a), uniformly rotating configurations $(\hat{A}^{-1}=0.0)$ can satisfy the PSR~J0740$+$6620 mass constraint only at high central densities, near the stability boundary. 
%As differential rotation is progressively applied
As the degree of differential rotation increases, the maximum mass increases and equilibrium configurations satisfying the observed mass become accessible over a progressively wider range of central densities, extending to lower values. This indicates that differential rotation not only raises the maximum mass but also broadens the parameter space of viable neutron star configurations consistent with the observational constraint.

We find that the FSUGarnet+SU(6) EoS (Fig.~\ref{fig:PSR}b), even with differential rotation applied, fails to support a neutron star mass consistent with the observed value of $2.14^{+0.10}_{-0.09}\,M_{\odot}$, as the hyperonic degrees of freedom substantially soften the EoS beyond what differential rotation can compensate. This softening is inherent to the SU(6) symmetric coupling scheme, which provides the conservative treatment of hyperonic degrees of freedom among hyperon coupling schemes. Since SU(6) symmetry is known to be broken in nature, more refined approaches such as SU(3)-based models or phenomenological hyperon-nucleon interactions could produce a stiffer hyperonic EoS~\cite{Miyatsu2025}. Given that differential rotation can substantially increase the maximum mass, a broader investigation of hyperon-nucleon couplings remains both meaningful and necessary for resolving the hyperon puzzle.

%--------------------------------------------------------------------
\begin{figure}[t]
    \centering
    \includegraphics[width=\columnwidth]{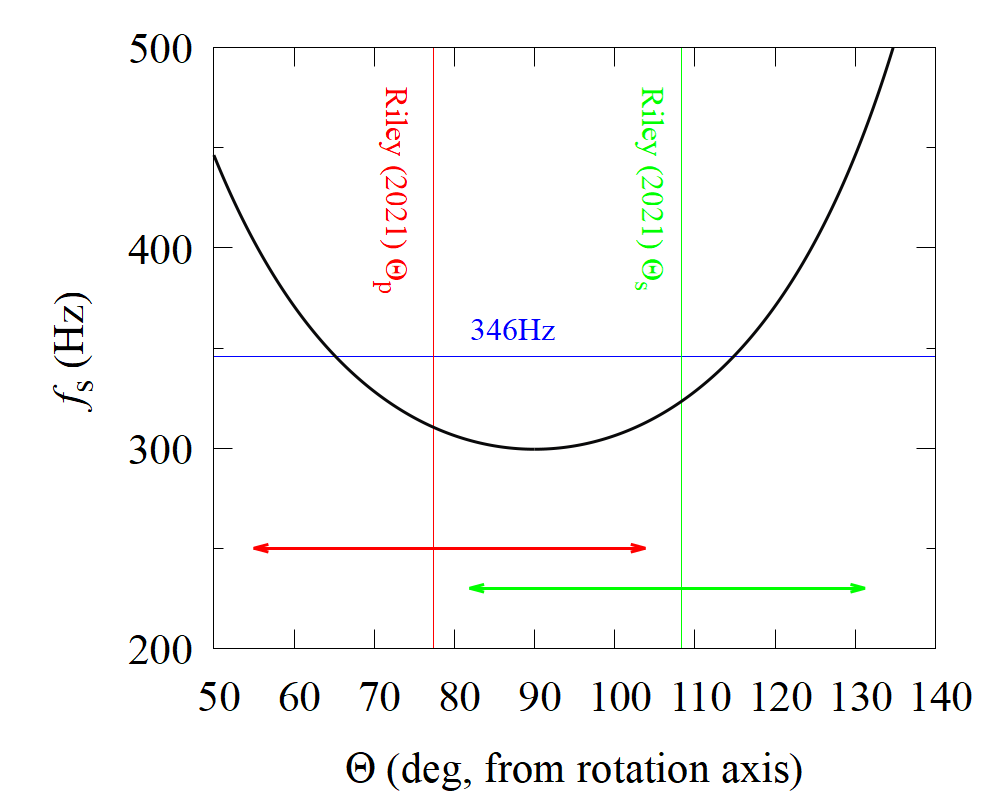}\vspace{-5mm}
    \caption{Surface spin frequency $f_s$ as a function of the angle
            $\Theta$ measured from the rotation axis, for a selected neutron
            star model with $f_e = 300$\,Hz, and $\hat{A}^{-1}=2.5$.}
    \label{fig:surface}
\end{figure}
%--------------------------------------------------------------------

We note that the analysis presented in this section is based on the
equatorial spin frequency. Several studies, however, suggest that the
hot spots of PSR~J0740+6620 are not located on the equator. While for
uniform rotation the spin frequency is identical everywhere on the
surface, differential rotation causes the surface frequency to vary
with the angle from the rotation axis. The analysis of Riley
\textit{et al.}~\cite{Riley2021} infers hot spots at colatitudes
$\Theta_p = 77.4^{+26.4}_{-22.4}$ deg and
$\Theta_s = 108.3^{+22.9}_{-26.4}$ deg. In Fig.~\ref{fig:surface}, we show
the angular dependence of the surface spin frequency for a selected
model with $f_e = 300$\,Hz and $\hat{A}^{-1}=2.5$. The surface frequency of this star does not exceed 346\,Hz at the equator, nor at the median hot-spot
colatitudes; however, if the hot spots are located farther from the
equator, the local surface frequency can exceed 346\,Hz. This implies
that equilibrium configurations with larger $\hat{A}^{-1}$ may remain
consistent with the observed spin frequency when the off-equatorial
hot-spot location is taken into account, which could in turn affect
the mass constraints inferred for PSR~J0740+6620. However, the current uncertainty in the hot-spot colatitudes is too large to draw a definitive conclusion.

\subsubsection{GW\,190814}

%--------------------------------------------------------------------
\begin{figure}[t]
    \centering
    \includegraphics[width=\columnwidth]{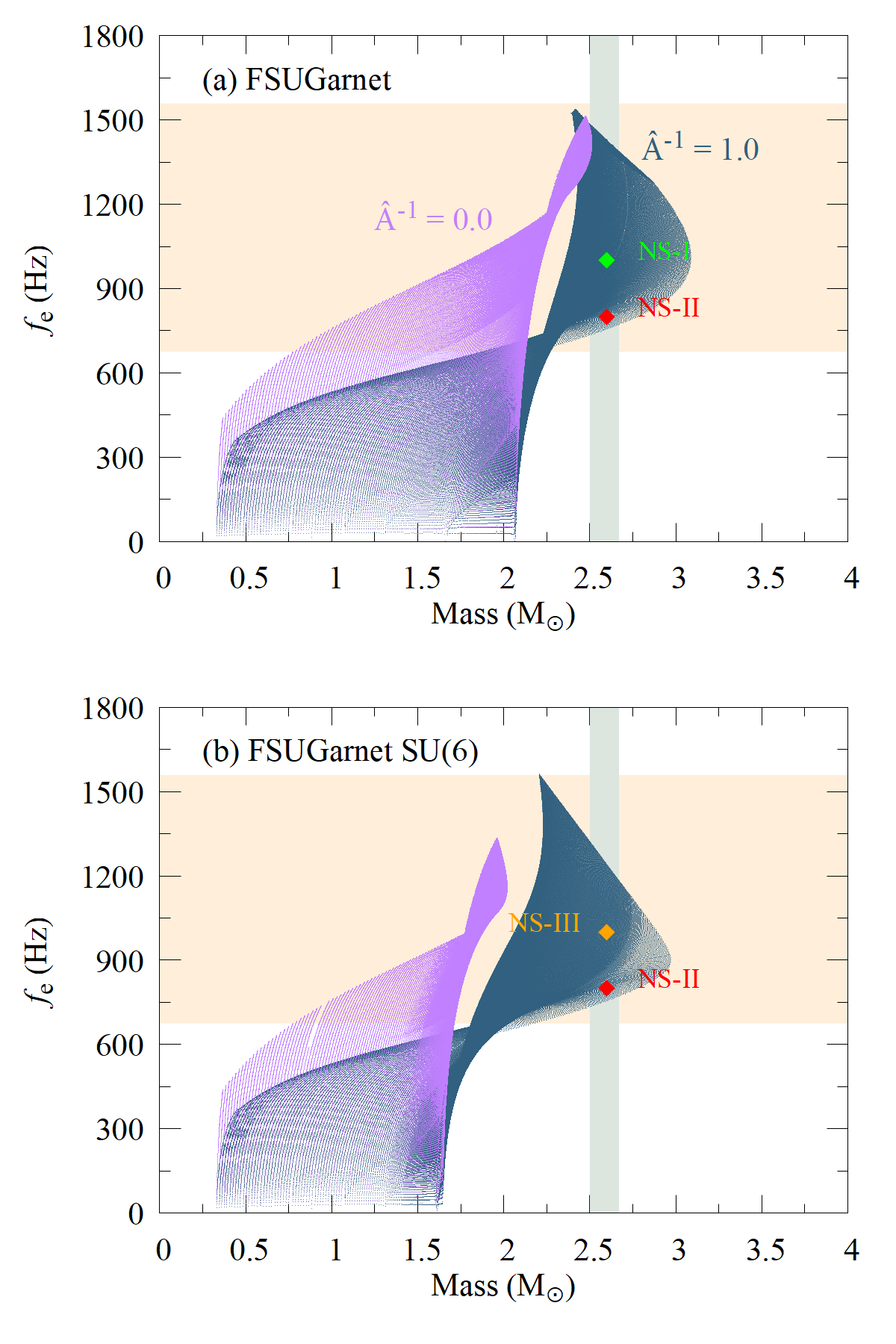}\vspace{-5mm}
    \caption{Equatorial spin frequency $f_e$ as a function of gravitational mass $M$ for all computed equilibrium configurations with a fixed differential rotation parameter $\hat{A}^{-1}=1.0$ (steelblue) and $\hat{A}^{-1}=0.0$ (purple), using the FSUGarnet (a) and FSUGarnet+SU(6) (b) EoSs. the vertical shaded band indicates the mass range of GW\,190814 (2.50 - 2.67 $M_\odot$), and the horizontal shaded band indicates the $90\%$ confidence interval for the spin frequency of the secondary component, $f=1170^{+389}_{-495}$ Hz~\cite{Biswas2021}.}
    \label{fig:gw}
\end{figure}
%--------------------------------------------------------------------

GW\,190814 is a compact binary coalescence event detected via 
gravitational waves, in which the secondary object has an inferred 
mass of $2.50$--$2.67\,M_\odot$. This places it in an ambiguous 
regime between a massive neutron star and a light black hole, making 
it one of the most debated objects in compact star physics. Unlike for 
pulsars, the spin frequency of the secondary component is not directly 
observed, and long-term stability of differential rotation need not 
be imposed; it suffices to consider equilibrium and stability only 
on the merger timescale. In particular, a Bayesian analysis by 
Biswas \textit{et al.}~\cite{Biswas2021} suggests that if the secondary object 
is a rapidly rotating neutron star, its spin frequency is constrained 
at the 90\% confidence level to $f = 1170^{+389}_{-495}$\,Hz. We note that this analysis assumes a uniformly rotating neutron star; allowing for differential rotation could shift the inferred spin frequency.

Figure~\ref{fig:gw} shows all equilibrium configurations obtained with a fixed differential rotation parameter $\hat{A}^{-1}=0$ and $1$, by varying the central density from $\rho_c = 3.25 \times 10^{14} \, \text{g} \, \text{cm}^{-3}$  to  $2.32 \times 10^{15} \, \text{g} \, \text{cm}^{-3}$ and the axis ratio from $r_{\text{ratio}}=1.0$ down to $0.3$ in steps of $0.001$. This value $\hat{A}^{-1}=1$ is chosen as a representative case, as other values of $\hat{A}^{-1}$ yield qualitatively similar trends, differing only in the accessible ranges of mass and rotational frequency.

We first examine the case of uniform rotation (purple), corresponding to the same rotation law assumed in the spin frequency constraint $f=1170^{+389}_{-495}$ Hz. For FSUGarnet, equilibrium configurations satisfying the GW\,190814 mass constraint could be obtained at sufficiently high spin frequencies. In contrast, for FSUGarnet+SU(6), no such equilibrium configuration could be found even at high rotation rates, since the maximum mass of the non-rotating (static) configuration is already too low. This result, together with those shown in Fig.~\ref{fig:tov}, indicates that while nucleonic matter alone can satisfy the GW\,190814 constraint once rotation is taken into account, the inclusion of hyperonic degrees of freedom makes this impossible—revealing yet another facet of the hyperon puzzle.

On the other hand, when differential rotation is included, both EoS 
models can reproduce equilibrium configurations satisfying the 
observed mass. For FSUGarnet, a substantially larger fraction of the 
scanned parameter space (41306 out of 280799 configurations, i.e., 
14.7\%) satisfies the mass constraint over a wide range of spin 
frequencies. FSUGarnet+SU(6), on the other hand, satisfies the mass 
constraint only over a more restricted range of spin frequencies, 
yielding a smaller fraction (15066 out of 280799, i.e., 5.4\%) than 
FSUGarnet. In other words, under the same scanned parameter range, a 
substantially larger fraction of the FSUGarnet+SU(6) models fail to 
reach the observed mass.

%--------------------------------------------------------------------------
\begin{table}[t]
\caption{Properties of the three representative differentially rotating neutron star configurations as shown in Fig.~\ref{fig:gw}. NS-I and NS-II are computed with the FSUGarnet EoS, while NS-III is computed with the FSUGarnet+SU(6) EoS.
}\vspace{2mm}
\begin{tabular}{lccccc}
\hline\hline
Neutron Star & $r_{\text{ratio}}$ & $\rho_c$ $( 10^{15} \,\text{g}/\text{cm}^3)$ & $R_e$ (km) & $f_e$ (Hz) & $\beta$\\
\hline
NS-I & 0.517 & 1.33 & 13.2704 & 1000 & 0.149  \\
NS-II & 0.378 & 0.55 & 17.2473 & 800 & 0.211 \\
NS-III  & 0.385 & 0.97 & 14.8363 & 1000 & 0.205 \\
\hline\hline
\end{tabular}\label{tab:NS_GW}
\end{table}
%--------------------------------------------------------------------------

%--------------------------------------------------------------------------
\begin{figure}[t]
    \centering
    \includegraphics[width=\columnwidth]{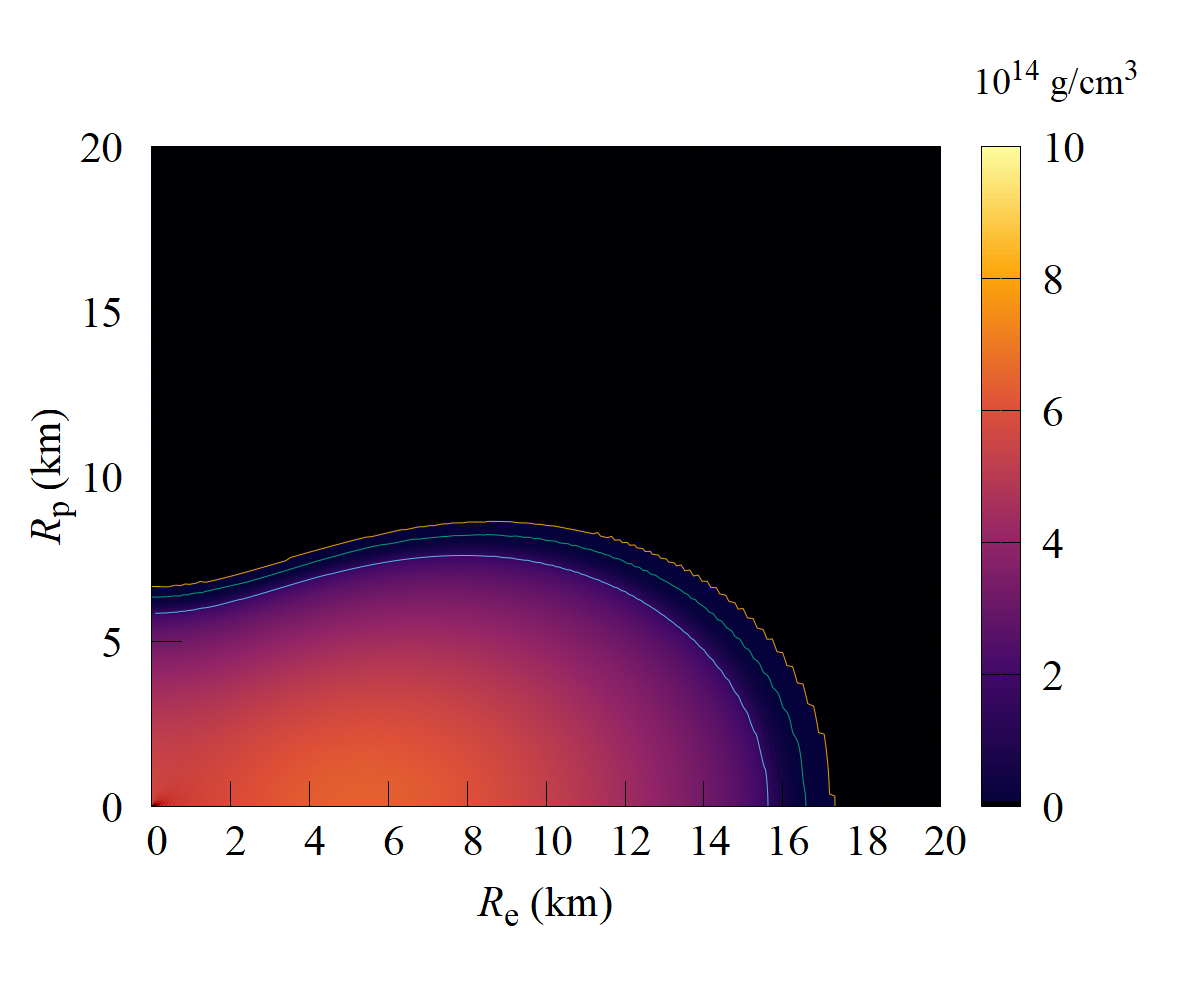}\vspace{-0.5mm}
    \caption{Two-dimensional density distribution of NS-II in the meridional plane, shown as a function of the equatorial radius $R_e$ and polar radius $R_p$. The color scale indicates the mass density in units of $10^{14} \, \text{g}/\text{cm}^3$. Contour lines indicate the boundaries between the outer crust, inner crust, and outer core.}
    \label{fig:gwden1}
\end{figure}
%--------------------------------------------------------------------------

For uniformly rotating neutron stars, rotation primarily modifies the radial density profile, while the overall trends and internal structure remain qualitatively similar to those of static neutron stars. We therefore focus instead on the internal structures that can emerge specifically under differential rotation.

%--------------------------------------------------------------------------
\begin{figure*}[h!]
    \centering
    \includegraphics[width=0.90\textwidth]{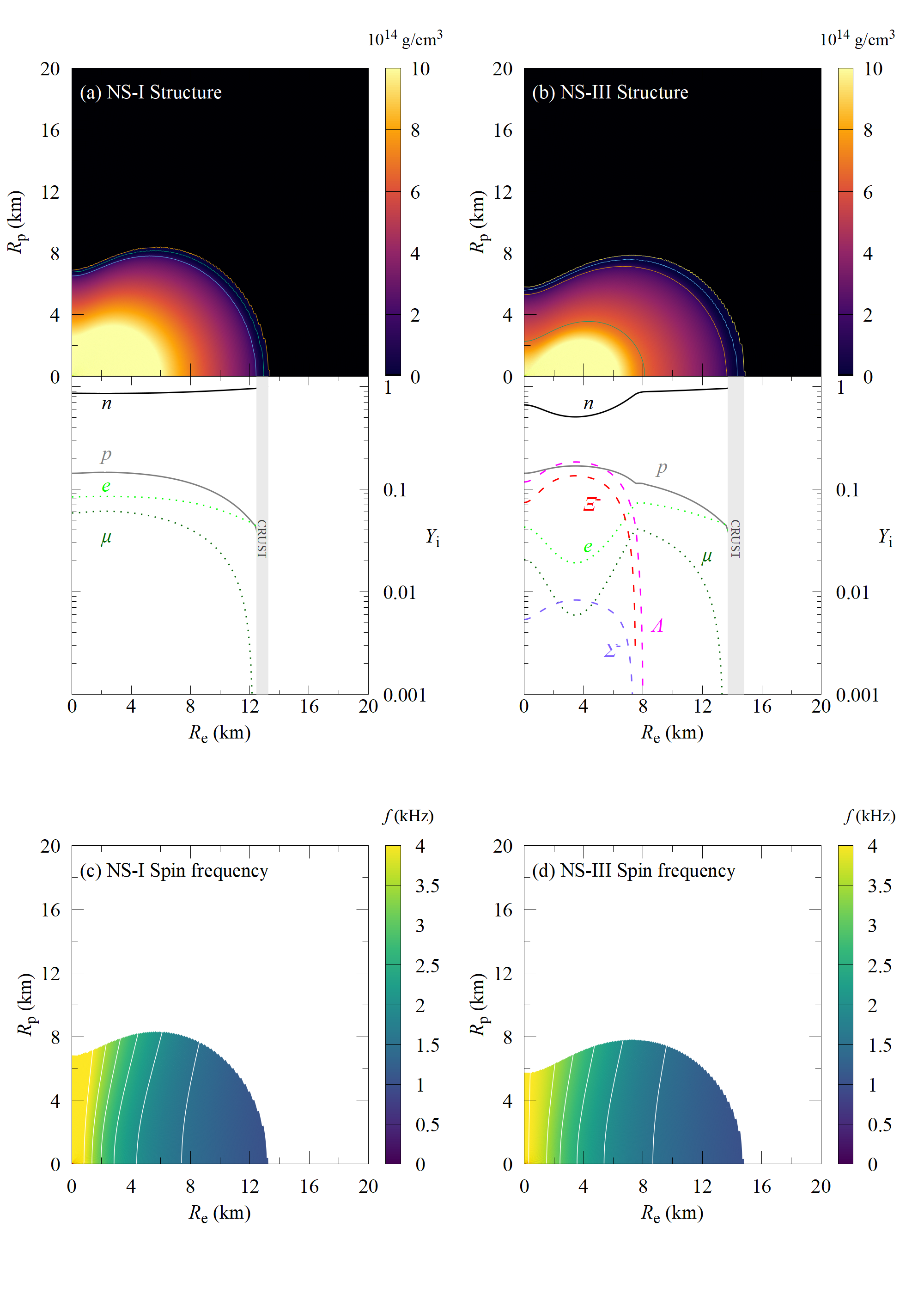}
    \vspace{-6ex}
    \caption{Internal structure and spin frequency distributions of differentially rotating neutron stars. Upper panels show the density profiles (in units of $10^{14}~\mathrm{g~cm^{-3}}$) for (a) NS-I and (b) NS-III configurations, along with the particle fraction $Y_i$ as a function of equatorial radius $R_e$. 
    In NS-I, the core composition includes nucleons ($n$, $p$), electrons ($e$), and muons ($\mu$), while NS-III additionally contains hyperons ($\Lambda$, $\Sigma$, $\Xi$). The gray shaded region indicates the crust-core boundary. 
    Lower panels display the local spin frequency $f$ (in kHz) distributions 
    for (c) NS-I and (d) NS-III, with white contour lines denoting isosurfaces of constant frequency.}
    \label{fig:gwcom}
\end{figure*}
%--------------------------------------------------------------------------
\clearpage

To examine the internal structure of differentially rotating neutron stars satisfying the GW\,190814 constraints, we select three representative configurations from Fig.~\ref{fig:gw}, all with a gravitational mass of $2.6 \, M_\odot$. NS-I and NS-II are both computed with the FSUGarnet EoS but differ in their equatorial spin frequency, while NS-I and NS-III share the same spin frequency but are computed with different EoSs (FSUGarnet and FSUGarnet+SU(6), respectively). The detailed properties of each configuration are listed in Table~\ref{tab:NS_GW}.

Figure~\ref{fig:gwden1} shows the two-dimensional density distribution of NS-II. Interestingly, NS-I and NS-III exhibit distinct stellar structures despite sharing the same mass and spin frequency, owing to the appearance of hyperons at high central densities. NS-II, on the other hand, achieves equilibrium at relatively low central densities for the given mass and spin frequency, which remain below the hyperon onset density. Consequently, the hyperonic degrees of freedom are never activated, and FSUGarnet and FSUGarnet+SU(6) yield identical configurations for NS-II.

Figure~\ref{fig:gwcom} presents a detailed analysis of the internal structure and properties of NS-I and NS-III. In the FSUGarnet case (Fig.~\ref{fig:gwcom}a), the absence of hyperons yields a stiffer EoS, so that equilibrium configurations satisfying $M=2.6M_\odot$ and $f_e=1000 \, \text{Hz}$ require only moderate deformation, with $r_{\text{ratio}}=0.517$ and $\beta=0.149$. In contrast, the FSUGarnet+SU(6) case (Fig.~\ref{fig:gwcom}b) provides a softer EoS due to the presence of hyperons, necessitating significantly stronger deformation to achieve the same mass and spin frequency, with $r_\text{ratio}=0.378$ and $\beta = 0.205$. Notably, the high-density interior of NS-III contains regions where $\Lambda$, $\Xi^-$, and $\Sigma^-$ hyperons appear sequentially. However, the central density is insufficient to reach the onset densities of the heavier hyperons $\Xi^0$, $\Sigma^0$, and $\Sigma^+$.

A notable feature of differentially rotating neutron stars, in contrast to static or uniformly rotating configurations, is that the density maximum is displaced from the rotation axis, occurring at a small offset from the center, as clearly seen in Fig.~\ref{fig:gwcom}. This is a consequence of the differential rotation law, which imparts higher angular velocity to the central region than to the outer layers, effectively pushing the central material outward.

Figures~\ref{fig:gwcom}c and \ref{fig:gwcom}d show the local spin frequency distributions inside NS-I and NS-III, respectively. While the equatorial surface frequency is $f_e=1000 \, \text{Hz}$ for both configurations, the frequency increases toward the center, reaching up to $ \sim 4000 \, \text{Hz}$ in the core region. The white contour lines represent isosurfaces of constant angular velocity, drawn at intervals of 500\,Hz. For the rotation law adopted in this work [Eq.~\eqref{eq:rotlaw}], the angular velocity decreases approximately quadratically with the distance from the rotation axis, so that the spacing between successive contour lines widens outward.

A direct assessment of the dynamical stability of NS-I, NS-II, and NS-III would require numerical simulations. However, \citet{Shibata2000b} demonstrated that the parameter $\beta$ serves as a reliable diagnostic for the onset of dynamical instability in differentially rotating relativistic neutron stars. In their study, configurations with $\hat{A}^{-1}$ values comparable to those considered here were found to become unstable at $\beta \gtrsim 0.24$. Since all three selected configurations have $\beta$ values below this threshold, they are expected to be dynamically stable on the dynamical timescale.

\subsection{Neutron stars in extreme conditions}

Differential rotation allows for the construction of more extreme, exotic neutron star equilibrium configurations than uniform rotation. In this section, we explore the extreme equilibrium configurations accessible with the FSUGarnet+SU(6) EoS.

\subsubsection{High-density configuration}

\begin{figure}[h]
    \centering
    \includegraphics[width=\columnwidth]{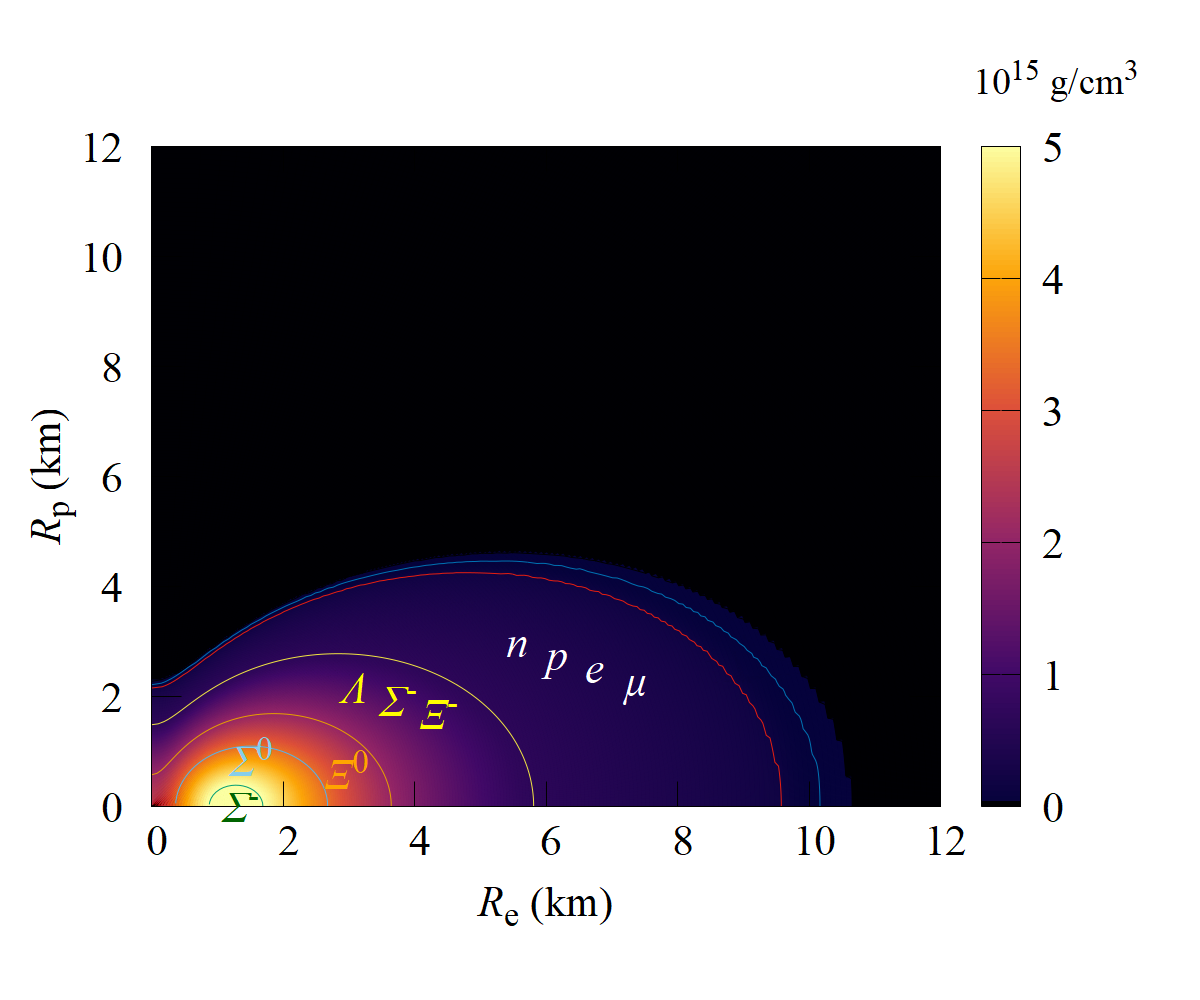}\vspace{-5mm}
    \caption{Two-dimensional density distribution of an extreme differentially rotating neutron star configuration with $\rho_c = 2.5 \times 10^{15} \, \text{g}/\text{cm}^3$, $r_\text{ratio}=0.2$, and $\hat{A}^{-1}=1$, computed with the FSUGarnet+SU(6) EoS. The color scale indicates the mass density in units of $10^{15} \, \text{g}/\text{cm}^3$. Contour lines delineate the regions where each baryon species becomes populated, with labels indicating the composition in each zone.}
    \label{fig:hi}
\end{figure}

Under strong differential rotation, the density in the off-center region can reach values two to three times higher than the central density. This gives rise to extremely high-density interiors exceeding ten times the nuclear saturation density, which are generally difficult to realize in other configurations, and enables the appearance of heavier hyperons. An example of such a configuration is presented in Fig.~\ref{fig:hi}.

%--------------------------------------------------------------------------
%--------------------------------------------------------------------------

Figure~\ref{fig:hi} shows an extreme configuration with a central density of $\rho_c = 2.5 \times 10^{15} \, \text{g}/\text{cm}^3$, an axis ratio of $r_\text{ratio}=0.2$, and a differential rotation parameter of $\hat{A}^{-1}=1$, yielding a gravitational mass of $M=2.19 \, M_\odot$ and an equatorial radius of $R_e = 11.26 \, \text{km}$. In this configuration, the maximum density exceeds $5 \times 10^{15} \, \text{g}/\text{cm}^3$ in the off-center high-density region, producing a density distribution sufficient for all eight members of the baryon octet to appear. The equatorial surface frequency is $f_e = 1763 \, \text{Hz}$, while the central rotation frequency reaches $9211 \, \text{Hz}$, corresponding to an extreme ratio of approximately five. The value of $\beta=0.26$ exceeds the instability threshold discussed above, and this configuration is therefore expected to be dynamically unstable on the dynamical timescale, although it serves as an illustration of the extreme density regimes accessible under strong differential rotation.

A more refined hyperonic EoS could yield different hyperon onset densities and appearance sequences, potentially allowing for dynamically stable extreme configurations within an adjusted parameter space.

%--------------------------------------------------------------------------
\begin{figure}[h]
    \centering
    \includegraphics[width=\columnwidth]{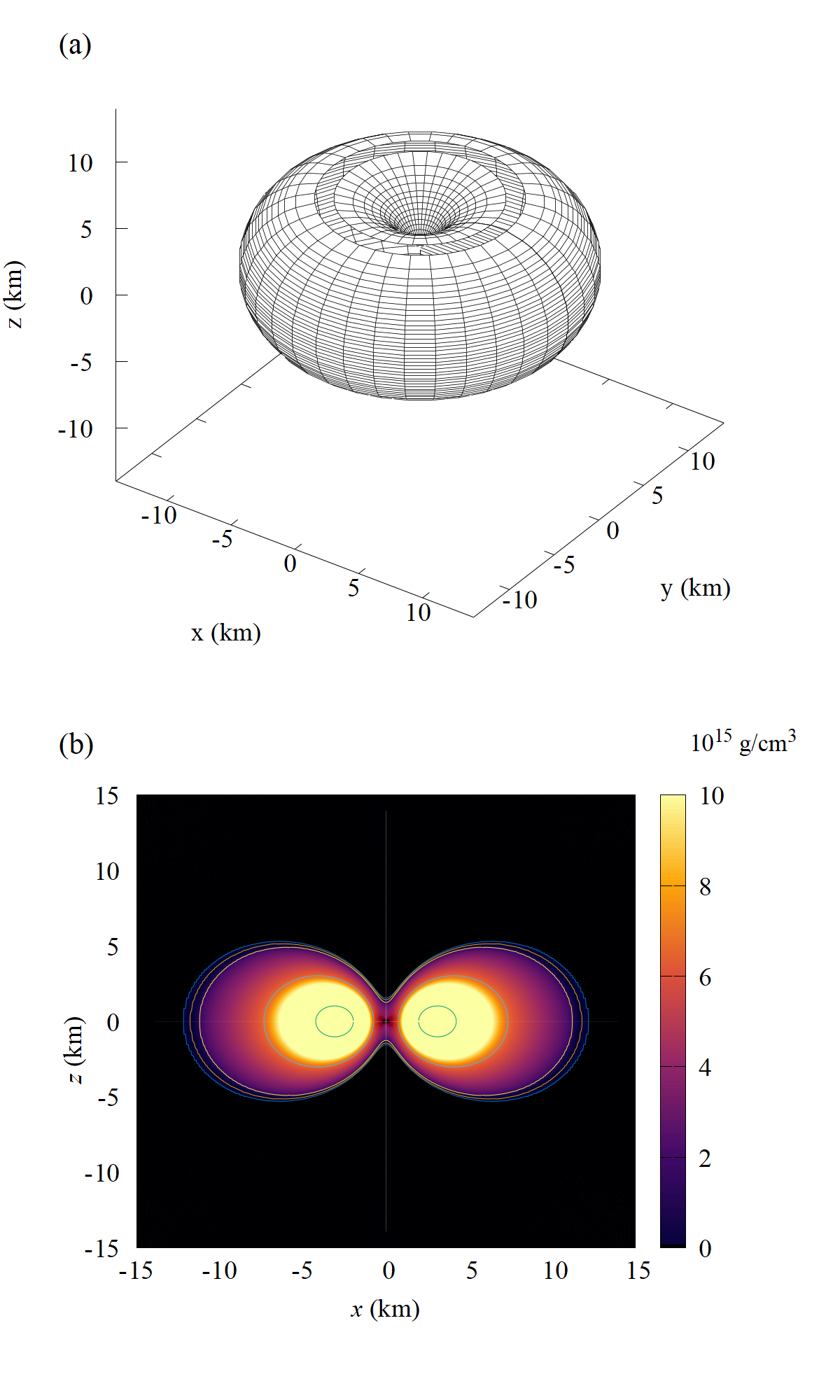}\vspace{-5mm}
    \caption{(a) Three-dimensional surface representation of a representative quasi-toroidal neutron star configuration. (b) Density distribution in the $x-z$ plane for the same configuration. The innermost green contour marks the onset of $\Xi^0$ hyperons, while the sky blue contour marks the onset of $\Lambda$, $\Sigma^-$, and $\Xi^-$ hyperons.}
    \label{fig:toroidal}
\end{figure}
%--------------------------------------------------------------------------

\subsubsection{Quasi-toroidal configuration}

Another intriguing extreme configuration is the \textit{quasi-toroidal} neutron star. Under sufficiently strong differential rotation, the stellar shape evolves progressively from an oblate spheroid toward a quasi-toroidal shape, where the shape looks toroidal, but there remains a non-vanishing density at the center. Figure~\ref{fig:toroidal} shows the three-dimensional plot and density profile of such an extreme configuration.

This configuration is computed with a central density of $\rho_c = 4 \times 10^{14} \, \text{g}/\text{cm}^3$ and a differential rotation parameter of $\hat{A}^{-1}=1.0$. The gravitational mass is $M=2.98M_\odot$ and the equatorial surface frequency is $f_e = 1326 \, \text{Hz}$. Despite the low central density, the off-center high-density region exceeds the hyperon onset density owing to differential rotation. The value of $\beta = 0.29$ exceeds the secular instability threshold of $\beta_{\text{sec}} \approx 0.14$ and even surpasses the dynamical instability threshold of $\beta_{\text{dyn}} \approx 0.27$ found for oblate spheroids, indicating that this configuration is significantly more dynamically unstable than the sample configurations analyzed previously.

Although maintaining dynamical stability in a quasi-toroidal neutron star is likely challenging, one can envision a scenario in which such a configuration forms transiently following a binary neutron star merger, when differential rotation is expected to be strongest. As angular momentum is subsequently redistributed (\textit{e.g.}, via magnetic braking, viscosity, or gravitational-wave emission) differential rotation gradually weakens, and the stellar shape evolves from a quasi-toroidal shape toward a more spheroidal configuration, eventually settling into an HMNS supported primarily by differential rotation. As differential rotation continues to decay, such a configuration is ultimately expected to collapse to a black hole once rotational support becomes insufficient to sustain its mass.

If such a scenario is realized, Fig.~\ref{fig:toroidal} suggests the intriguing possibility of a \textit{hyperon ring}: a toroidal region where hyperons are present at off-center radii, while the low-density core remains composed of ordinary nuclear matter without hyperonic degrees of freedom. Stability and dynamical evolution of such exotic configurations require further careful investigations.

\section{Summary and prospect}\label{Sec:Conclusion}

We have investigated the effects of differential rotation on neutron star structure and properties using two realistic nuclear equations of state (EoSs): FSUGarnet, which includes only nucleonic degrees of freedom, and its hyperonic extension FSUGarnet+SU(6). Through this study, we have examined how differential rotation affects key stellar properties, particularly the gravitational mass and rotational frequency.

First, we have performed calculations at a fixed spin frequency of 346\,Hz, corresponding to the observed rotation rate of PSR~J0740+6620, to assess whether differential rotation provides a viable scenario for resolving the hyperon puzzle, in which the inclusion of hyperonic degrees of freedom softens the EoS and reduces the maximum mass, making it difficult to sustain observed massive neutron stars exceeding $2M_\odot$. We have found that differential rotation can substantially increase the maximum mass; however, 346\,Hz alone is insufficient to produce a significant mass enhancement, and the FSUGarnet+SU(6) EoS proved too soft to reproduce the observed mass of PSR~J0740+6620. Nevertheless, as discussed above, the SU(6) symmetric coupling scheme represents a conservative choice, and FSUGarnet was not originally constructed with hyperonic properties in mind. Given the considerable uncertainties in the EoS at supranuclear densities, a refined hyperonic EoS stiffer than FSUGarnet+SU(6) remains a realistic possibility, in which case differential rotation could serve as a viable mechanism for resolving the hyperon puzzle.

Next, we have examined the internal structures of neutron stars consistent with the observational constraints from GW\,190814. A comparison between NS-I, which contains only nucleonic degrees of freedom, and NS-III, which incorporates hyperonic degrees of freedom, reveals that the stiffer FSUGarnet EoS requires less deformation to achieve the same equatorial surface frequency and gravitational mass. In contrast, NS-II demonstrates that equilibrium configurations satisfying the observational constraints are achievable even at relatively low central densities; in this case, the maximum density does not reach the hyperon onset density, and both FSUGarnet and FSUGarnet+SU(6) yield identical internal structures. For NS-III, where hyperons are present, the density maximum is displaced from the center, giving rise to a spatially varying particle fraction profile in the interior. All three configurations have $\beta$ values below 0.24, making them viable candidates for the secondary component of GW\,190814, although a detailed investigation of their stability on the dynamical timescale remains necessary.

Finally, we have investigated neutron stars under extreme conditions. Under strong differential rotation with significant deformation, the off-center density can reach values two to three times higher than the central density. Under appropriate conditions, this enables the formation of density regimes sufficient for all hyperons in the baryon octet to appear simultaneously. At extreme conditions, we have found that neutron stars can form quasi-toroidal structures, in which hyperons are distributed in a toroidal region — what we call a \textit{hyperon ring} — surrounding a core of ordinary nuclear matter. While such extreme configurations are unlikely to be dynamically stable on the dynamical timescale, they may exist transiently as hypermassive neutron stars (HMNSs) in the immediate aftermath of a binary neutron star merger.

Through this study, we identify several important implications. First, the uncertainties in the EoS at supranuclear densities play a crucial role in the context of differentially rotating neutron stars. The FSUGarnet parameter set used in this work was designed to barely satisfy the $\approx 2.0 M_\odot$ constraint of PSR~J0740+6620, and its extension via SU(6) symmetry, while representing an extreme choice, ultimately failed to reproduce the observed mass. A modest adjustment of the symmetry energy and incompressibility within the ranges permitted by current terrestrial accelerator experiments would be sufficient to construct a stiffer EoS. If further refined to reproduce hypernuclear properties based on SU(3) symmetry, such an EoS could plausibly accommodate the observational constraints under differential rotation, even with hyperonic degrees of freedom included. A more refined EoS is therefore essential for providing a more accurate picture of neutron star structure in these extreme regimes.

This study also reaffirms the utility of the Komatsu-Eriguchi-Hachisu (KEH) method. The KEH method reformulates the Einstein field equations as a set of Poisson-like integral equations, providing a numerically stable and computationally efficient approach. Given that computational resources remain limited, performing full general relativistic hydrodynamics (GRHD) simulations for all possible scenarios is practically infeasible. The KEH method, on the other hand, allows for the computation of a large number of equilibrium configurations, as demonstrated in this work. By evaluating $\beta$ and other relevant quantities across a wide parameter space, it provides an efficient means of identifying promising candidates for targeted GRHD simulations. Furthermore, the KEH method goes beyond simply providing initial conditions such as mass and radius; it enables the construction of equilibrium configurations with diverse internal matter distributions under identical macroscopic conditions, making it a valuable tool for investigating the internal structure of neutron stars in the context of merger simulations. As a natural next step, we plan to employ the equilibrium configurations constructed here as initial data for GRHD evolutions with the GPU-accelerated code AthenaK~\cite{Stone2026}, in order to directly assess the dynamical stability and evolution of the configurations identified in this work.

The rotation law adopted in this work represents a simplified prescription, as the actual rotational profile of neutron stars remains poorly constrained. Several attempts have been made to explore alternative or modified rotation laws~\cite{zhou2019,Galeazzi2012,Cassing2024}. Since the internal structure of a neutron star is divided into the crust and core, with different particle compositions at each density regime, determining which rotation law is physically appropriate and sustainable requires a parallel investigation of both the rotational profile and the internal matter composition. Such studies will be essential for constructing more realistic models of differentially rotating neutron stars.

In summary, this work demonstrates that differential rotation provides a viable framework for exploring massive neutron stars with hyperonic degrees of freedom. More refined EoSs, improved rotation laws, and efficient numerical methods such as the KEH scheme will be essential for fully resolving the hyperon puzzle and characterizing the extreme configurations accessible in neutron star mergers.

\section*{Acknowledgments}
H. K. would like to thank Jorge Piekarewicz and Tsuyoshi Miyatsu for valuable advice on constructing EoSs from relativistic mean-field theory, and Young-Min Kim and Gyeongbin Park for helpful discussions on neutron stars. This work was supported by JSPS Grant-in-Aid for Scientific Research, Grants No. JP23K03410, No. JP23K25864, No. JP25H01269, and No. JP26KJ1140. This work was also supported by Sasakawa Scientific Research Grant from the Japan Science Society. J. K. acknowledges support from the Basic Science Research Program through the National Research Foundation of Korea (NRF) funded by the Ministry of Education (RS-2021-NR065815). We thank APCTP, Pohang, Korea for their hospitality during the programs APCTP-2026-S09 and APCTP-2026-T01 from which this work greatly benefited.

\bibliography{reference}

\end{document}